\documentclass[aps,prl,twocolumn,showpacs,10pt,preprintnumbers,nofootinbib,reprint]{revtex4-1}
\usepackage[margin=7pt,justification=centerlast]{caption}

\usepackage{amsmath}
\usepackage{amsfonts}
\usepackage{amssymb}
\usepackage{graphicx, rotating}
\usepackage{epstopdf}
\usepackage{epsfig}
\usepackage{latexsym}
\usepackage{mathtools}
\usepackage{graphicx}
\usepackage{color}
\usepackage[dvipsnames]{xcolor}
\usepackage{amsmath,amssymb}
\usepackage{multirow}
\usepackage{array,makecell}
\usepackage[utf8]{inputenc}
\usepackage{ mathrsfs }
\usepackage{tabularx}
\usepackage{ bbold }

\usepackage{slashed}
\usepackage{hyperref}
\hypersetup{colorlinks, citecolor=bluscuro, linkcolor=bluscuro, urlcolor=bluscuro}
\definecolor{rossos}{cmyk}{0,1,1,0.55}
\definecolor{bluscuro}{rgb}{0.15, 0.2, .85}
\definecolor{bluchiaro}{cmyk}{1,.3,0.,0.1}
\definecolor{verdescuro}{rgb}{0.3,0.8,0.3}

\newcommand{\be}{\begin{equation}}
\newcommand{\ee}{\end{equation}}          
\newcommand{\bea}{\begin{eqnarray}}
\newcommand{\eea}{\end{eqnarray}}
\newcommand{\bc}{\begin{center}}
	\newcommand{\ec}{\end{center}}

\begin{document}

\title{On the extraction of $\alpha_\textit{em}(m_Z^2)$ at Tera-$Z$}

\preprint{CERN-TH-2025-005}

\author{Marc Riembau}
\affiliation{Theoretical Physics Department, CERN, 1211 Geneva 23, Switzerland}

\begin{abstract}
\noindent 
{
The current projected sensitivity on the electromagnetic coupling  $\alpha_\textit{em}(m_Z^2)$ represents a bottleneck for the precision electroweak program at FCC-ee.
We propose a novel methodology to extract this coupling directly from $Z$-pole data. 
By comparing the differential distribution of electrons, muons and positrons in the forward region, the approach achieves a projected statistical sensitivity below the $10^{-5}$ level, representing a significant improvement over other methods.
We assess the impact of leading parametric uncertainties including that of the  top quark mass.
}

\end{abstract}

\maketitle

\section{Introduction}

The $Z$-pole run and its potential production of $6\cdot 10^{12}$ $Z$ bosons is emerging as the flagship of the future FCC-ee program \cite{FCC:2018evy}. However, the near \textit{per-million} statistical precision on some observables is useless unless accompanied by a herculean effort to bring theoretical and experimental uncertainties to a comparable level, 
and to design well-motivated observables.
A replica of the LEP program is insufficient.
For instance, the $10^{-4}$ relative uncertainty on the electromagnetic coupling at the $Z$ pole $\alpha_\textit{em}(m_Z^2)$ 
is a major obstacle for obtaining accurate Standard Model (SM) predictions.

The current uncertainty on $\alpha_\text{em}(m_Z^2)$ is dominated by the hadronic contribution $\Delta\alpha_\text{had}^{(5)}(m_Z^2)$ to the running from its low energy measurement \cite{atoms7010028}, while the leptonic contribution is known at four loops \cite{Steinhauser:1998rq,Sturm:2013uka}. Different approaches for the hadronic contribution lead to 
$\Delta\alpha_\text{had}^{(5)}(m_Z^2)\times 10^4=276.1\pm 1.0$
\cite{Erler:2017knj}, 
$275.23\pm 1.2$
\cite{Jegerlehner:2019lxt,Proceedings:2019vxr}, 
$276.0\pm 1.0$ \cite{Davier:2019can}, 
$276.1\pm 1.1$ 
\cite{Keshavarzi:2019abf}, 
the lattice value $277.3\pm 1.5$ \cite{Ce:2022eix}, and the PDG average $278.3\pm 0.6$ \cite{ParticleDataGroup:2024cfk}.
The main sources of uncertainty come from the $e^+e^-\to\text{hadrons}$ cross section below $\sqrt{s}<2\,\text{GeV}$ and un-calculated higher order perturbative and non-perturbative corrections, see  \cite{Jegerlehner:2019lxt,Davier:2019can}. In the future, bringing the cross section measurement below the $1\%$ level and the perturbative calculation below the $0.1\%$ level translate to $4\cdot 10^{-4}$ uncertainty on $\Delta\alpha_\text{had}^{(5)}(m_Z^2)$ \cite{Jegerlehner:2019lxt}. 
Since the electroweak precision program at FCC-ee requires a relative precision on $\alpha_\text{em}$  of $10^{-5}$, equivalent to an absolute precision of $10^{-4}$ on $\Delta\alpha_\text{had}^{(5)}(m_Z^2)$, and given potential tensions between data and lattice results \cite{Davier:2023cyp,Erler:2024lds}, an alternative, complementary, and direct extraction of $\alpha(m_Z^2)$ is highly desirable. 

A proposal to extract $\alpha(m_Z^2)$ directly at FCC-ee is found in Ref.~\cite{Janot:2015gjr}, based on measuring the forward-backward asymmetry of muon production, $A_{FB}^{\mu\mu}$, during off-peak runs at $\sqrt{s_-}=87.9\,\text{GeV}$ and $\sqrt{s_+}=94.3\,\text{GeV}$ and comparing it with the one measured on-peak. At $\sqrt{s_\pm}$, $A_{FB}^{\mu\mu}$ depends on both $\alpha(m_Z^2)$ and the effective mixing angle $\sin^2\theta_W^\textit{eff}$, whereas at the $Z$-pole it is sensitive only to $\sin^2\theta_W^\textit{eff}$. The measurement of $\alpha(m_Z^2)$ is statistics limited, with an expected relative uncertainty of $3\cdot 10^{-5}$, equivalent to a $4\cdot 10^{-4}$ absolute uncertainty on $\Delta\alpha_\text{had}^{(5)}(m_Z^2)$, comparable to the $10^{-5}$ expected relative sensitivity on $\sin^2\theta_W^\textit{eff}$. 
Additionally, as we discuss in this work, the quoted precision leverages the top mass  accuracy that will be obtainable after the $t\bar{t}$ threshold run at FCC-ee.

 \begin{figure}
	\centering
	\includegraphics[width=1\linewidth]{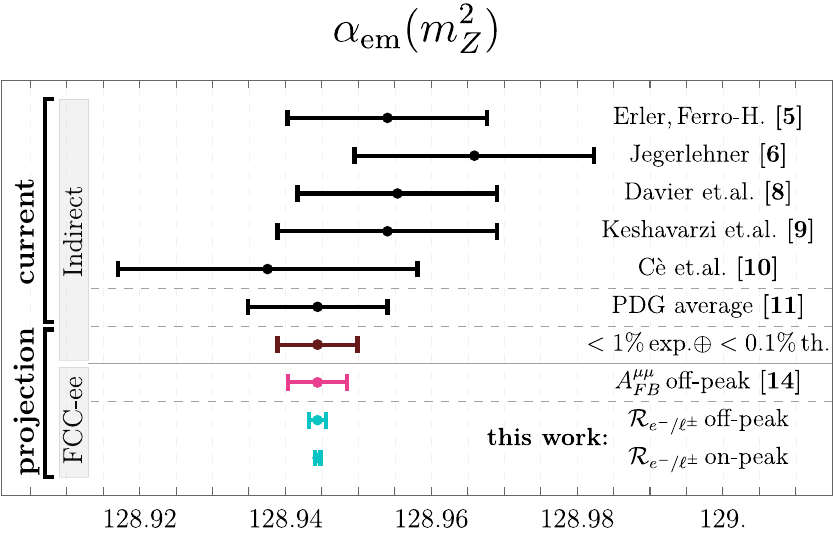}
\end{figure}

In this \textit{Letter} we present a novel method to extract $\alpha(m_Z^2)$ directly from $Z$ pole measurements, 
which presents a significant improvement in statistical sensitivity compared to other methods. The production rates of electrons as a function of the scattering angle is compared with those of muons and of positrons. In the forward region, 
for angles $\theta\lesssim 30^\circ$ but still well within the detector acceptance, 
muon and positron production are driven by the $Z$ pole exchange, while electron production receives an equally sizable contribution from the forward photon pole. We demonstrate that the proposed observables reach a statistical sensitivity on $\alpha(m_Z^2)$ below the $10^{-5}$ level.

\section{$Z$-pole sensitivity on $\alpha_\textit{em}$ and $\sin^2\theta_W^\textit{eff}$}

A dominant process in $e^+e^-$ colliders is the electroweak Bhabha scattering, $e^+e^-\to e^+e^-+X$, where $X$ represents soft and collinear emissions, and has been extensively studied at LEP \cite{Beenakker:1990mb,BEENAKKER1998199,Placzek:1999xc,Montagna:1999tf}. 
At very small scattering angles the process is dominated by QED and can be used to monitor the collider luminosity \cite{JADACH1995349,CarloniCalame:2015zev}. At FCC-ee, the cross section measurement between 62 and 88 mrad is expected to allow the determination of the absolute luminosity at the $10^{-4}$ level \cite{Jadach:2018jjo,Dam:2021sdj}. 

At intermediate scattering angles, above 100~mrad and well within the detector, the dominant contributions to the forward electroweak Bhabha scattering arise from the forward $t$-channel photon pole and the $Z$ $s$-channel pole, which are of comparable size and statistically significant. The former is enhanced by the forward photon pole, while the latter is enhanced by a $\frac{m_Z^2}{\Gamma_Z^2}$ factor. Since the $Z$-exchange leads to an imaginary amplitude, while the photon exchange is real, their interference vanishes and the process is dominated by the individual squared amplitudes. 
At leading order, the contribution from the photon $t$-channel pole at $\sqrt{s}=m_Z^2$ is proportional to $\frac{\alpha^2}{4m_Z^2}\frac{2((1+c_\theta)^2+4)}{(1-c_\theta)^2}$, while the $Z$ exchange is proportional to
$\frac{\alpha^2}{4m_Z^2}\frac{m_Z^2}{\Gamma_Z^2}\mathcal{Z}^2(1+c_\theta^2+8c_\theta r_Vr_A)$, where $c_\theta$ is the scattering angle and we have defined $\mathcal{Z} = \frac{\sqrt{2}G_F m_Z^2}{\pi \alpha}(g_V^2+g_A^2)$, with $g_V=\frac12T^3_e-Q_e\sin^2\theta_W^\textit{eff}$ and $g_A = \frac12T^3_e$, and $r_{V,A}=g^2_{V,A}/(g_V^2+g_A^2)$ \cite{CONSOLI1979208,Caffo:367852,Altarelli:1989hv}. 
Important loop effects affecting the running coupling in the $t$-channel will be discussed later in detail. 
Writing $z\equiv \frac{1-c_\theta}{2}$ and approximating $r_Vr_A\simeq 0$, at leading order in $z\ll 1$ the ratio between the two contributions is of order one at $\frac12z^2\simeq \left(\frac{m_Z}{\Gamma_Z}\mathcal{Z}\right)^{-2}$, which corresponds to $c_\theta\simeq 0.8$ or $35^\circ$. Consequently this suggests that an accurate measurement of electroweak Bhabha scattering at the $Z$ pole for angles $c_\theta \gtrsim 0.8$ would be sensitive to the overall parameter $\mathcal{Z}$, which depends on both the electromagnetic coupling $\alpha$ and the effective mixing angle $\sin^2\theta_W^\text{eff}$. 

We define two different observables sensitive to $\alpha_\text{em}$ and  $\sin^2\theta_W^\text{eff}$, and independent of the absolute luminosity normalization. First, the ratio between the number of electrons and the number of muons produced at a fixed angle $\theta$, $\mathcal{R}_{e^-/\mu^-}(\theta)$. Second, the ratio between the number of electrons and the number of positrons produced at a fixed angle, $\mathcal{R}_{e^-/e^+}(\theta)$. While statistically independent\footnote{Note that the ratio $\mathcal{R}_{e^-/\mu^+}(\theta)$ is equivalent to measuring $\mathcal{R}_{e^-/\mu^-}(\theta)$ and the muon forward-backward asymmetry $A_{FB}^{\mu\mu}$ if CP is assumed.}, these two ratios probe similar physics. As argued, for scattering angles $c_\theta\gtrsim 0.8$, electron production is sensitive to both the $Z$ pole and the photon pole, while muon and positron production occur predominantly through the $Z$-boson exchange. At larger angles, the contribution from the photon pole becomes negligible and $\mathcal{R}_{e^-/\mu^-}(\theta)\to 1$. Instead, measurement of  $\mathcal{R}_{e^-/e^+}(\theta)$ at larger angles becomes equivalent to the measurement of $A_{FB}^{ee}$.

\smallskip

\textit{---Statistical power.} \quad Assessing the expected statistical uncertainty on the observables $\mathcal{R}_{e^-/\ell^\pm}$ at Tera-$Z$ and the corresponding constraint on  $\alpha_\text{em}$ and  $\sin^2\theta_W^\text{eff}$ is important as it provides a target for the rest of the uncertainties and the ultimate reach.

The ratio $\mathcal{R}_{e^-/\ell^\pm}(\theta)$ is assumed to be measured for $c_\theta\geq 0$ in bins $c_\theta\in[\theta_i,\theta_{i+1}]$ of uniform size $\theta_{i+1}-\theta_i=0.05$, except the most forward bin which only includes $c_\theta\in[0.95,0.99]$. By cutting at $c_\theta=0.99$ we therefore assume the detector to cover up to $\theta\simeq 140\,\text{mrad}$, consistent with the planned coverage up to $120\,\text{mrad}$ \cite{Barchetta:2021ibt}. We comment later the impact of reducing the effective detector coverage.  
The statistical uncertainty in each bin, denoted $\delta\mathcal{R}_{e^-/\ell^\pm}^i$, is given by $\delta\mathcal{R}_{e^-/\ell^\pm}^i = \mathcal{R}_{e^-/\ell^\pm}^i \sqrt{N_{e^-}^{-1}+N_{\ell^\pm}^{-1}}$, where $N_{e^-}(N_{\ell^\pm})$ is the number of electrons (positrons/muons) at each bin $i$, with $c_\theta\in[\theta_i,\theta_{i+1}]$, and we have assumed no correlation.

Given our focus on the estimate of statistical uncertainties, the impact of NLO corrections is minimal and we use the tree level estimation for the total rates. 
The full NLO electroweak corrections to Bhabha scattering have been known for a long time \cite{CONSOLI1979208,BERENDS1983537,CAFFO1985378,Tobimatsu:1985vd,Tobimatsu:1985pp,BOHM1988687}, while at two loops, only the log-enhanced 2 loop corrections are known \cite{Kuhn:2001hz,Feucht:2004rp,Jantzen:2005az,Penin:2011aa}. The situation for the pure QED case is more advanced, as the full two loop contribution is known for massless fermions \cite{Bern:2000ie}, with the massive case computed in \cite{Penin:2005eh,Mitov:2006xs,Becher:2007cu,Actis:2007gi,Bonciani:2004gi,Czakon:2006pa,Actis:2009uq,Henn:2013woa,Duhr:2021fhk,Delto:2023kqv} and the two loop hadronic in \cite{Actis:2007fs,Kuhn:2008zs}. We checked that the full NLO corrections for the SM $e^+e^-\to e^+e^-$ differential cross section, as computed with \verb|ReneSANCe|	\cite{Sadykov:2020any,Bondarenko:2022mbi}, give a correction to the rates, dominated by the large QED logs, with negligible changes in the statistical reach. 
The study of  $\delta\mathcal{R}_{e^-/\ell^\pm}$ at higher orders in perturbation theory, and the extent of the cancellations in the ratio and the requirements to reduce the theoretical uncertainty below the statistical sensitivity is left for future work. Due to the similarity of the processes, we expect an effort comparable to the MUonE initiative, which targets a precision of $10^{-5}$ for the $\mu e\to\mu e$ process \cite{Banerjee:2020tdt,Banerjee:2020rww,Banerjee:2021mty,Broggio:2022htr}. 
In the present work, we focus on the loop effects that bring new parametric uncertainties, which will be addressed in detail in subsequent discussions.

\begin{figure}
	\centering
	\includegraphics[width=1\linewidth]{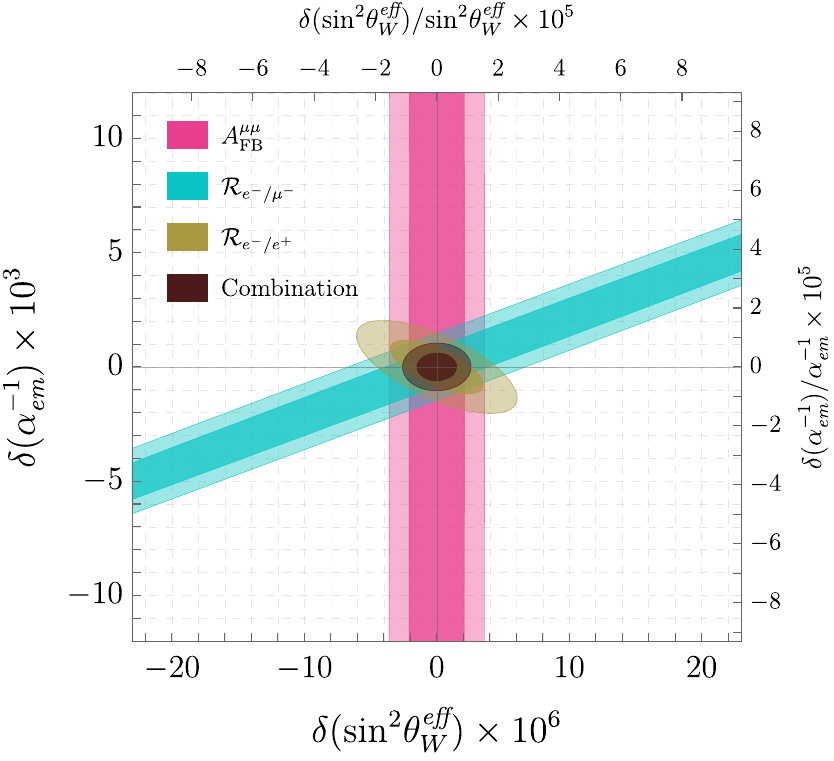}
	\caption{One and two sigma expected statistical uncertainties on $\alpha^{-1}_\text{em}$ and $\sin^2\theta_W^\text{eff}$ from the muon forward-backward asymmetry $A^{\mu\mu}_{FB}$ (pink), the electron to muon ratio $\mathcal{R}_{e^-/\mu^-}$ (teal)  and the electron to positron ratio $\mathcal{R}_{e^-/e^+}$ (gold), from 125/ab at $\sqrt{s}=m_Z$.}
	\label{fig:sineffaemAFBvsRATstatonly}
\end{figure}

The target luminosity for the Tera-$Z$ run at $\sqrt{s}=m_Z$ is of $140\times10^{34}/\text{cm}^2\text{s}$ per interaction point  \cite{janot_2024_yr3v6-dgh16}. Following the reference, assuming $1.2\cdot 10^7$ effective second per year and four interaction points leads to $68/\text{ab}$ per year and around $10^{11}$ $Z\to \mu\mu$ decays. The target for the total integrated luminosity at $\sqrt{s}=m_Z$ is 125/ab. This leads to a relative statistical uncertainty of $10^{-5}$ in the last bin $[0.95,0.99]$, and rising almost linearly up to $2\cdot 10^{-5}$ in the central bins. 
The ratio in each bin is interpreted as a measurement of $\alpha_\text{em}$ and  $\sin^2\theta_W^\text{eff}$. By writing $\alpha^{-1}_\text{em}=128.964+\delta(\alpha^{-1}_\text{em})$ and $\sin^2\theta_W^\text{eff}=0.23148+\delta(\sin^2\theta_W^\text{eff})$, the statistical reach with 125/ab is shown in Fig.~\ref{fig:sineffaemAFBvsRATstatonly}. The measurement of $\delta\mathcal{R}_{e^-/\mu^-}$ constrains a specific direction in the $\delta(\alpha^{-1}_\text{em})-\delta(\sin^2\theta_W^\text{eff})$ plane at a relative precision of $\sim 10^{-5}$, as shown in teal in Fig.~\ref{fig:sineffaemAFBvsRATstatonly}. This direction is almost orthogonal to $\delta(\sin^2\theta_W^\text{eff})$, which is constrained through the muon forward-backward asymmetry $A^{\mu\mu}_{FB}$ at a similar relative precision, shown in pink in Fig.~\ref{fig:sineffaemAFBvsRATstatonly}.
The measurement of $\delta\mathcal{R}_{e^-/e^+}$ at small angles constrains a direction in the plane, but at large angles it becomes a measurement of $A_{FB}^{ee}$ and therefore sensitive to $\delta(\sin^2\theta_W^\text{eff})$ directly, resulting in the ellipse in Fig.~\ref{fig:sineffaemAFBvsRATstatonly}.
This allows for the combination to have a statistical sensitivity to both $\delta(\alpha^{-1}_\text{em})$ and $\delta(\sin^2\theta_W^\text{eff})$ at the $10^{-5}$ level.

Given that the number of electrons scales as $1/(1-c_\theta)^2$ in the forward region while the number of muons and positrons is roughly constant, the statistical sensitivity on the ratios $\mathcal{R}_{e^-/\ell^\pm}$ is controlled by the muon and positron cross section. Therefore, the reach has small dependence on the complete detector coverage as long as the region around $c_\theta\simeq 0.8$ is well under control. As mentioned, the projected sensitivity in Fig.~\ref{fig:sineffaemAFBvsRATstatonly} assumes to saturate the statistical precision up to $c_\theta=0.99$, corresponding to angles of $\theta\simeq 8^\circ$ or $120\,\text{mrad}$, and reaches a combined relative sensitivity on $\alpha_\text{em}$ of $0.6\cdot 10^{-5}$. Assuming to cover instead only up to $c_\theta=0.98$ ($\theta\simeq 11^\circ$ or $200\,\text{mrad}$) has negligible impact on the reach, and coverage up to $c_\theta=0.95$ ($\theta\simeq 18^\circ$ or $320\,\text{mrad}$) leads to a mild effect with a relative statistical precision on $\alpha_\text{em}$ of $0.7\cdot 10^{-5}$. Coverage up to $c_\theta= 0.85$ ($\theta\simeq 32^\circ$) leads instead to an order one effect with a relative reach on $\alpha_\text{em}$ of $1.5\cdot 10^{-5}$.

\smallskip

\textit{--- Systematic uncertainties.}\quad  
A first source of systematic uncertainties is given by the particle miss-identification rate. Miss-id between electrons and muons were already below the $10^{-5}$ level for the ALEPH detector at LEP \cite{ALEPH:1994ayc}. In order for an event to contribute, it requires a double miss-id, and therefore we assume that this effect will be well under control with FCC-ee detectors.

Charge miss-identification is at the 0.5\% level at LEP \cite{ALEPH:1999smx}. 
In order for an event to contribute to either $\mathcal{R}_{e^-/e^+}$ or $\mathcal{R}_{e^-/\mu^-}$, it requires a double charge miss-id. As long as FCC-ee detectors provide charge id better than $\sim 0.2\%$ in the region $\theta\lesssim 20^\circ$, this leaves a negligible effect as well. In comparison, the measurement of $A_{FB}^{\mu\mu}$ does require a similar level of control on charge-id \cite{Janot:2015gjr}, but does not rely as heavily as $\mathcal{R}_{e^-/\ell^\pm}$ on the forward region, where this requirement might prove more challenging. Given the $\sim 10^8\, Z\to \mu^\pm \mu^\pm $ and $e^\pm e^\pm$ at FCC-ee, charge miss-id will be measured with precision.

Another source of uncertainty is the possibility that the identification efficiency has a dependence on the polar angle $\theta$. In the case of electrons and positrons, if this efficiency is independent of the lepton charge then the ratio $\mathcal{R}_{e^-/e^+}$ is insensitive to this effect as it affects equally electrons and positrons. 
The situation is similar to $A_{FB}^{\mu\mu}$ \cite{Janot:2015gjr}. 
In the case of the ratio $\mathcal{R}_{e^-/\mu^-}$, it is not realistic to assume the same angular dependence for the electron and muon efficiencies, and therefore the measurement of this ratio might potentially receive large systematic uncertainties and make the $\alpha(m_Z^2)$ extraction from $\mathcal{R}_{e^-/\mu^-}$ unfeasible. 
It should be noticed however that only a ratio of efficiencies with the same angular dependence as the one induced by $\alpha(m_Z^2)$ on $\mathcal{R}_{e^-/\mu^-}(\theta)$ is degenerate with $\alpha(m_Z^2)$. 
It is to be explored whether an unbinned analysis with constrained functional forms for the efficiencies might lead to a competitive measurement of $\alpha(m_Z^2)$. 

The beam energy spread $\delta\sqrt{s}$ has a small effect $\delta\sqrt{s}/m_Z^2$ for the photon exchange diagram, but a large effect $\delta\sqrt{s}/\Gamma_Z^2$ for the $s$-channel $Z$ exchange. However, as found in \cite{Blondel:2019jmp}, by measuring the longitudinal boost in $\mu^+\mu^-$ events at the $Z$ pole the energy spread can be measured at the \textit{per-mille} level every four minutes. The impact is the same as in \cite{Janot:2015gjr} for $A_{FB}^{\mu\mu}$, concluding that monitoring the energy spread leads to a negligible uncertainty for $\mathcal{R}_{e^-/\ell^\pm}$.

\section{Parametric uncertainties}

We discuss the impact of parametric uncertainties, defined as those arising from input parameters whose precise determination is required for accurate predictions of the ratios $\mathcal{R}_{e/\ell}$ and the asymmetry $A_{FB}^{\mu\mu}$ within the SM. The Fermi constant, known with a precision better than $10^{-6}$ \cite{MuLan:2010shf}, and the $Z$ boson mass, expected to be measured at a similar level at FCC-ee \cite{FCC:2018evy}, have negligible impact. We find however important sources of parametric uncertainties that require detailed consideration, namely the running of $\alpha_\textit{em}$, the $Z$ width and the top mass.

\smallskip

\textit{--- The running of $\alpha_\textit{em}$.}\quad 
A conceptually very important effect that arises at one-loop level is the correction to photon's propagation due to matter, absorbed in the running coupling. At a given scale $s$, the running coupling is given by $\alpha(s)=\alpha/(1-\Delta\alpha(s))$, with $\alpha$ being the electromagnetic coupling measured at zero momentum, known at the $10^{-10}$ level \cite{atoms7010028}, and $\Delta\alpha$ given by the vacuum polariation as detailed in, e.g., Ref.~\cite{Jegerlehner:2017gek}. In the $e^+e^-\to e^+e^-$ process, while the $s$-channel exchange depends on $\alpha_\text{em}(m_Z^2)$, the $t$-channel exchange, at a given scattering angle $c_\theta$, is sensitive to the running coupling $\alpha(t)$, evaluated at a momentum transfer $t=-\frac{m_Z^2}{2}(1-c_\theta)$. 
In order to interpret the measurements of $\mathcal{R}_{e^-/\ell^\pm}$ in terms of the $Z$-pole coupling $\alpha_\text{em}(m_Z^2)$, it is necessary to run the coupling between the two scales\footnote{The running of the electromagnetic coupling between different momentum transfers in Bhabha scattering was observed by the OPAL and L3 collaborations \cite{OPAL:2005xqs,L3:2005tsb}.}. At leading order in $\alpha_\text{em}$, this is given by
\be
\alpha(m_Z^2)\simeq \alpha(t) -\alpha\times \left(  \Delta\alpha(t)-\Delta\alpha(m_Z^2)  \right)\,.
\ee
Therefore, the accuracy on the $Z$-pole electromagnetic coupling $\alpha(m_Z^2)$ is limited solely by the accuracy on $\mathcal{R}_{e^-/\ell^\pm}$ only when $\Delta\alpha(t)-\Delta\alpha(m_Z^2)$ is of the same order than the statistical uncertainty on $\mathcal{R}_{e^-/\ell^\pm}$ at a given bin, around $10^{-5}$. Since the dominant uncertainty on $\Delta\alpha(t)-\Delta\alpha(m_Z^2)$ comes from the hadronic contributions to the vacuum polarization, we will focus our discussion on those. Writing $\Delta\alpha(t,m_Z^2)\equiv \Delta\alpha_\text{had}(t)-\Delta\alpha_\text{had}(m_Z^2)$, one has
\be
\Delta\alpha(t,m_Z^2) = \frac{\alpha}{3\pi}\int_{2m_\pi^2}^\infty \frac{ds}{s}R(s)\left( \frac{-t}{s-t}+\frac{m_Z^2}{s-m_Z^2} \right)
\ee
where $t<0$ and 
$R(s)$ is the so-called hadronic $R$-ratio. In order to evaluate it, we use the results of \cite{Davier:2017zfy,Davier:2019can} for the average of the $\pi^+\pi^-$, $\pi^+\pi^-\pi^0$, $2\pi^+2\pi^-$, $\pi^+\pi^-2\pi^0$ and $2K2\pi$ channels for $\sqrt{s}\leq 2\,\text{GeV}$ and the average for $e^+e^-\to\text{hadrons}$ between $3.7\,\text{GeV}$ and $5\,\text{GeV}$.
Between $2\,\text{GeV}$ and $3.7\,\text{GeV}$ and for $\sqrt{s}>5\,\text{GeV}$ we use instead the perturbative result, computed with the \verb|rhad| program \cite{Harlander:2002ur} which includes the perturbative calculation up to 4 loops. 
The uncertainties obtained for each individual channel coincide with those in \cite{Davier:2019can} when computing $\Delta\alpha_\text{had}^{(5)}(m_Z^2)$.

 \begin{figure}
	\centering
	\includegraphics[width=1\linewidth]{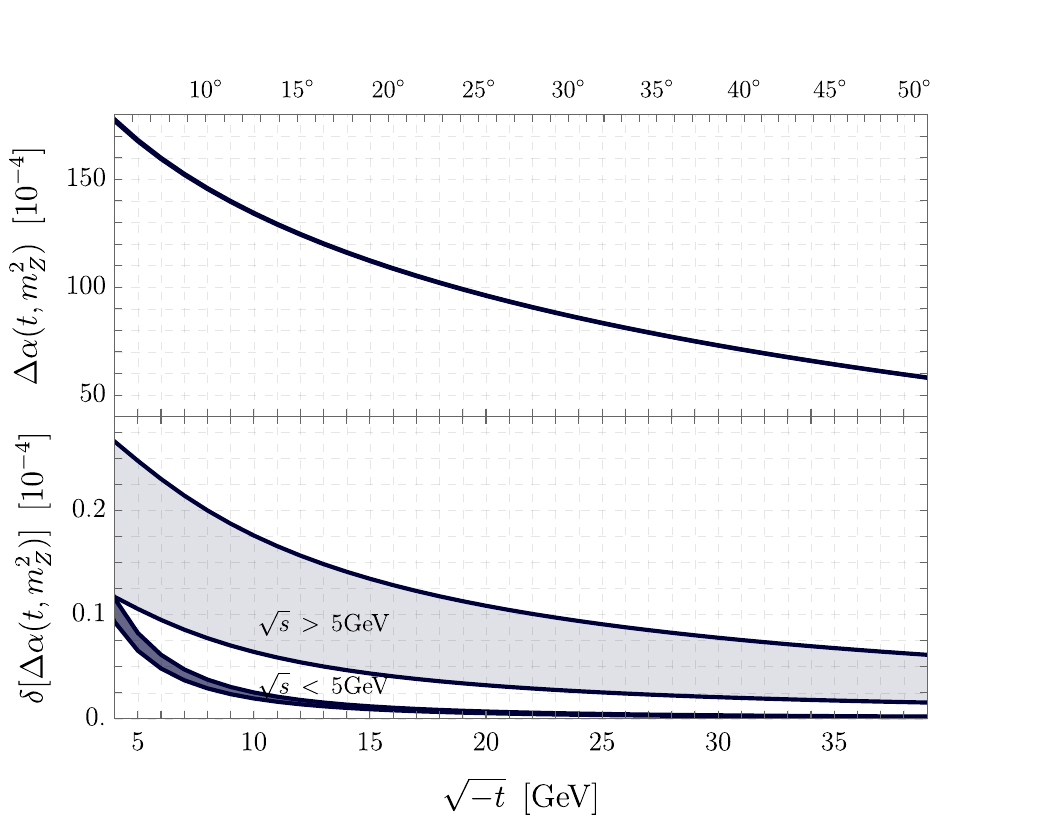}
	\caption{Top: $\Delta\alpha(t,m_Z^2)$, defined as $\Delta\alpha_\text{had}(t)-\Delta\alpha_\text{had}(m_Z^2)$, as a function of $t$. Bottom: Uncertainty on $\Delta\alpha(t,m_Z^2)$ coming from $R(s)$ for $\sqrt{s}<5\,\text{GeV}$ (darker) and from $\sqrt{s}>5\,\text{GeV}$ (lighter). See text for details.}
	\label{fig:deltatmz}
\end{figure}

The result for $\Delta\alpha(t,m_Z^2)$ as a function of the momentum transfer is shown in the upper plot of Fig.~\ref{fig:deltatmz}. It is strictly smaller than $\Delta\alpha_\text{had}(m_Z^2)\simeq 275\times 10^{-4}$ since the kernel $\frac{-t}{s-t}+\frac{m_Z^2}{s-m_Z^2}$ suppresses contributions for $s\ll |t|,m_Z^2$. In the lower plot we show the uncertainty on $\Delta\alpha(t,m_Z^2)$. 
In dark gray, the uncertainty coming from the $\sqrt{s}<5\,\text{GeV}$ in $R(s)$, dominated by the low energy $e^+e^-$ experiments in \cite{Davier:2017zfy,Davier:2019can}. The uncertainty is below the $10^{-5}$ level, since this region is highly suppressed due to the kernel. This is in contrast with the vacuum polarization contribution to the muon $g-2$ whose uncertainty is instead dominated by the two pion channel. The bulk of the uncertainty of $\Delta\alpha(t,m_Z^2)$ comes from perturbative QCD, which dominates the $\sqrt{s}>5\,\text{GeV}$ region, shown in lighter gray. The upper boundary of the band corresponds to the uncertainty obtained by computing $R(s)$ varying $m_c$, $m_b$ and $\alpha_S(m_Z)$ in the range $m_c=1.27\pm0.02\,\text{GeV}$, $m_b=4.18\pm0.03\,\text{GeV}$ and $\alpha_S(m_Z)=0.118\pm 0.0016$ \cite{Workman:2022ynf}, as well as evaluating the renormalization scale at $\mu=\sqrt{s}\times 2^{\pm1}$. The lower boundary of the band corresponds instead on varying only the renormalization scale $\mu$, while keeping fixed the other parameters. Given that the uncertainty is dominated by $\alpha_s(m_Z^2)$, which is expected to be significantly improved at FCC-ee, this represents a perfectly feasible scenario. We conclude therefore that the impact of the running uncertainty is at the $10^{-5}$ level in the most forward bin, while subdominant in the rest. This implies that there is no significant obstruction in interpreting the measurement of the ratio $\mathcal{R}_{e^-/\ell^\pm}$ in terms of $\alpha(m_Z^2)$.

\smallskip

\textit{--- The top mass and the $Z$ width.}\quad 
A second conceptually important effect that arises at one loop 
is the top contribution to the electroweak boson self-energy, leading to a parametric dependence on the top mass.
Taking as input parameters the Fermi constant $G_F$, the $Z$ boson mass $m_Z^2$ and the electromagnetic coupling $\alpha_\text{em}$, one has that the $Z$-boson gauge coupling to matter, given by $4\sqrt{2}G_F m_Z^2$, is corrected by the $T$ parameter due to the top-bottom mass splitting, $4\sqrt{2}G_F m_Z^2\to 4\sqrt{2}G_F m_Z^2\frac{1}{1-\Delta\rho}$ with $\Delta \rho = \frac{N_c \sqrt{2}G_F m_t^2}{16\pi^2}$ \cite{EINHORN1981146,CHANOWITZ1978285,Consoli:1989fg,Altarelli:1990zd,PhysRevD.46.381,Barbieri:2004qk}. 
Consequently, the top mass introduces a shift on the effective overall coupling, which can be written as $\delta\mathcal{Z}/\mathcal{Z} = 10^{-5}\times\frac{\delta m_t}{90\,\text{MeV}}$, implying that the top mass uncertainty is above the statistical sensitivity unless it is known at or below the $\sim~100\text{MeV}$ level. 
The projected $t\bar{t}$ threshold run of FCC-ee provides a constraint on $m_t$ at the $17\,\text{MeV}$ level, implying that the impact of the top mass on the $\alpha_\text{em}$ extraction is negligible once the complete set of FCC-ee data is considered. 

Since the $t\bar{t}$ run is scheduled after the Tera-$Z$ run, the initial interpretation of Tera-$Z$ data will likely rely on the top mass extracted from LHC data. Accordingly, we consider the scenario where Tera-$Z$ data is combined with the top mass obtained from HL-LHC measurements. 
The current uncertainty is well above the $100\text{MeV}$ level.  
The combined top quark mass measurement from ATLAS and CMS based on $\sqrt{s} = 7,8\,\text{TeV}$ data yields $m_t = 172.52 \pm 0.33\,\text{GeV}$ \cite{ATLAS:2024dxp}. 
However, nonperturbative effects introduce an additional $\sim 500\,\text{MeV}$ ambiguity in the interpretation of such measurements \cite{Nason:2017cxd,Azzi:2019yne,Hoang:2020iah,Dehnadi:2023msm}.
Such effects enter as well in the determinations of the top quark from $\sqrt{s}=13\,\text{TeV}$ data given by $171.17\pm0.38\,\text{GeV}$ from CMS \cite{CMS:2023ebf} and $174.41\pm0.8\,\text{GeV}$ from ATLAS \cite{ATLAS:2022jbw}, and the Tevatron combination of $174.30\pm0.65\,\text{GeV}$ \cite{CDF:2016vzt}. 
Theoretically cleaner measurements tend to be less accurate, e.g. the CMS measurement of $172.94\pm1.37\,\text{GeV}$ \cite{CMS:2022emx}. 
In the following we consider the uncertainty $\delta m_t=330\,\text{MeV}$ from the  $\sqrt{s}=7,8\,\text{TeV}$  combination as the figure of merit. This corresponds to a relative impact on $\delta \mathcal{Z}$ at the $\sim 4\cdot 10^{-5}$ level, potentially affecting the extraction of $\alpha_\text{em}$ from the ratio $R_{e^-/e^+}$ and the asymmetry $A_{FB}^{\mu\mu}$. 

\begin{figure}
	\centering
	\includegraphics[width=1\linewidth]{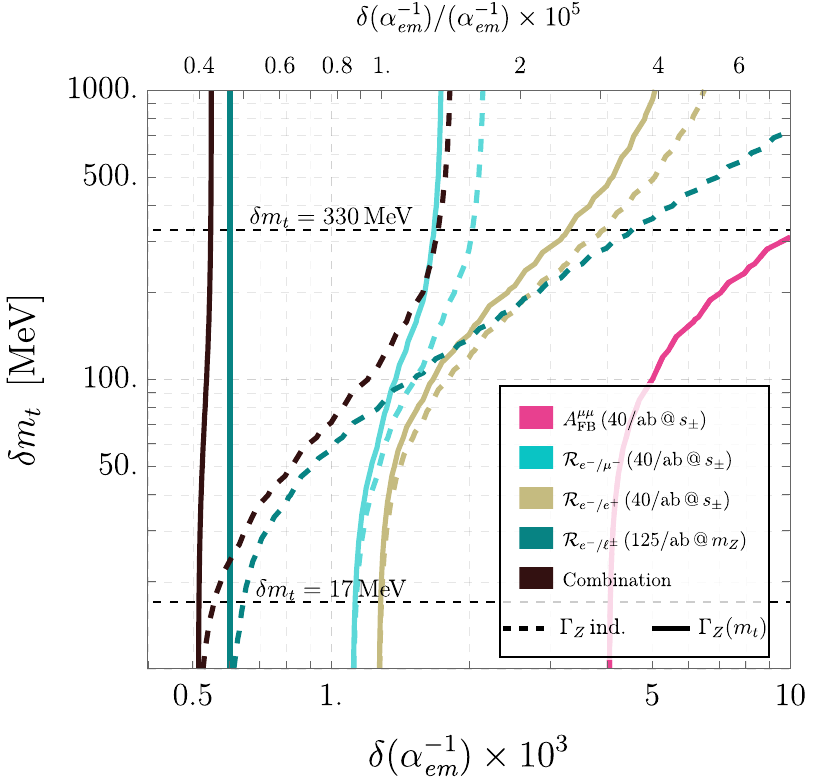}
	\caption{One-sigma expected statistical uncertainty on $\alpha_\text{em}^{-1}$ in the $A_{FB}^{\mu\mu}$, $\mathcal{R}_{e^-/\mu^-}$ and $\mathcal{R}_{e^-/e^+}$ observables as a function of the top mass uncertainty $\delta m_t$. See text for discussion on $\Gamma_Z$.}
	\label{fig:mtaem}
\end{figure}

The shift on the $Z$ coupling affects the $Z$-boson width in a similar manner, $\Gamma_Z\to \Gamma_Z\frac{1}{1-\Delta\rho}$. This implies that if the effect of the top on the electroweak self-energies is taken into account consistently, the $Z$-boson $s$-channel exchange at the $\sqrt{s}=m_Z$, proportional to $\mathcal{Z}/\Gamma_Z$ due to the resonant enhancement, is independent of $\Delta\rho$ and therefore has a reduced impact from the associated parametric uncertainty due to the top mass.
It is important to note that the current theoretical uncertainty on the $Z$-boson width, of $400\,\text{keV}$ \cite{Dubovyk:2018rlg}, is much larger than the expected experimental accuracy of $\delta\Gamma_Z\sim 11\,\text{keV}$ \cite{Blondel:2019jmp}.
While the effect of the width uncertainty on $A_{FB}^{\mu\mu}$ is suppressed, the effect on the ratio in the forward region is not, and one has that $\frac{\delta\mathcal{R}_{e^-/\ell^\pm}}{\mathcal{R}_{e^-/\ell^\pm}}\sim \frac{\delta\Gamma_Z}{\Gamma_Z}$. 
The current FCC-ee projection of $\delta\Gamma_Z\sim 11\,\text{keV}$ corresponds to $0.5\cdot 10^{-5}\times \Gamma_Z$, and therefore measurements of the ratio $\mathcal{R}_{e^-/\ell^\pm}$ have a sensitivity on the width comparable to the projected sensitivity from the line shape scan. This implies that the measurement from the line shape scan could be combined with the one extracted from $\mathcal{R}_{e^-/\ell^\pm}$.

The impact of incorporating the top mass dependence on the $\alpha_\text{em}$ extraction is summarized in Fig.~\ref{fig:mtaem}. We make use of two different treatments of the width.
First, we assume that the width is an independent parameter, fixed to some value obtained from the $Z$ line shape scan. This is indicated in dashed lines in Fig.~\ref{fig:mtaem}. Second, we assume that the theoretical calculation is improved and use this would-be prediction, with the leading shift proportional to $\Delta\rho$ canceling in the $s$-channel $Z$-boson exchange at $\sqrt{s}=m_Z$. We further assume in this plot that $\sin^2\theta_W^\textit{eff}$ is fixed to some value. A finite precision on the mixing angle has no qualitative effect on this plot since it is independently measured from $A_{FB}^{\mu\mu}$ at the $Z$ pole run, which has no sensitivity on $\alpha_\text{em}$.

We show in Fig.~\ref{fig:mtaem} the statistical sensitivity to $\alpha_\text{em}$ as a function of the top mass uncertainty $\delta m_t$. 
The FCC-ee projection on the top mass uncertainty of $17\,\text{MeV}$ leads to a determination of $\alpha_\text{em}$ equivalent to the $\delta m_t\to 0$ case for all the observables. 
Assuming a larger top mass uncertainty leads to a deterioration of the sensitivity in some cases. 
The off-peak measurements of $\mathcal{R}_{e^-/\mu^-}$ and $\mathcal{R}_{e^-/e^+}$, and of $A_{FB}^{\mu\mu}$, show a sensitivity to $m_t$. While the off-peak measurements of $\mathcal{R}_{e^-/\ell^\pm}$ present a mild dependence on the two treatments of the width discussed, $A_{FB}^{\mu\mu}$ has no dependence due to the reduced sensitivity on $\Gamma_Z$. 

The on-peak measurement of $\mathcal{R}_{e^-/\ell^\pm}$ shows a large dependence on the two treatments of the width. When the width is fixed, the top mass uncertainty through $\Delta\rho$ is correlated with $\alpha_\text{em}$ and the sensitivity is washed out. Only below $\delta m_t=100\,\text{MeV}$ the on-peak measurement competes with the off-peak ones. However, taking into account the shift on the width consistently, the on-peak measurement of $\mathcal{R}_{e^-/\ell^\pm}$ leads to an extraction of $\alpha_\text{em}(m_Z^2)$ robust against uncertainties associated to $m_t$.

\section{Sensitivity to the $S$ parameter}

\begin{figure}
	\centering
	\includegraphics[width=1\linewidth]{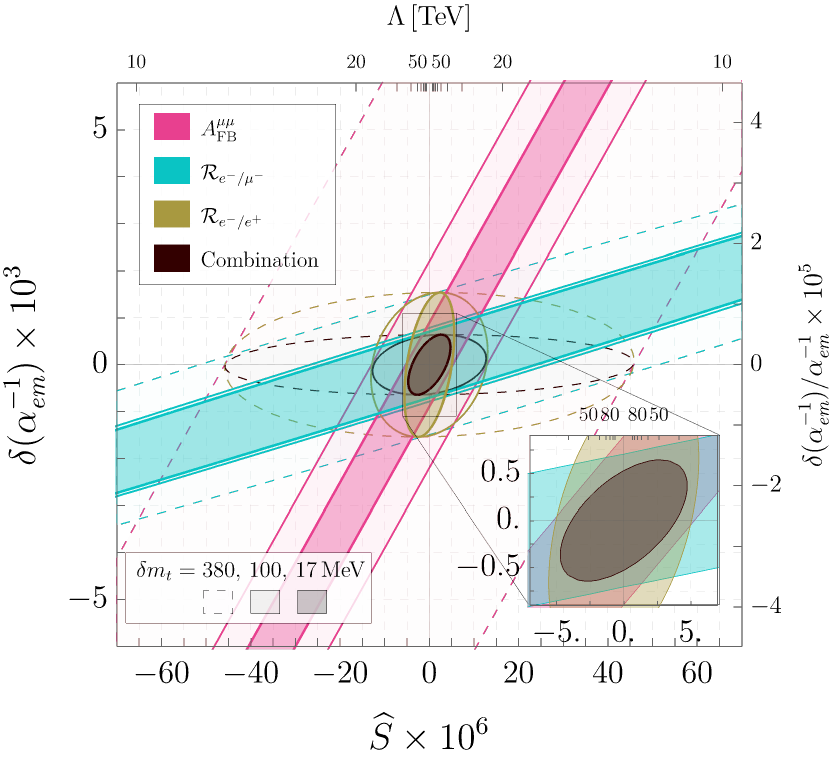}
	\caption{One sigma expected statistical sensitivity on $\alpha^{-1}_\text{em}$ and $\widehat{S}$ from the muon forward-backward asymmetry $A^{\mu\mu}_{FB}$ (pink), electron to muon ratio $\mathcal{R}_{e^-/\mu^-}$ (teal), and the electron to positron ratio $\mathcal{R}_{e^-/e^+}$ (gold), from 125/ab at $\sqrt{s}=m_Z$. Different shadings assume $\delta m_t=380, \,100,\text{ and } 17$~MeV. In the inset, only the FCC-ee projection $\delta m_t=17$~MeV is shown.}
	\label{fig:aem_Sh}
\end{figure}

The relevance of improving the extraction of $\alpha_\textit{em}$ is clear once we interpret the measurements in terms of a specific microscopic description. In the SM and in the scheme where $G_F$, $m_Z^2$ and $\alpha_\textit{em}$ are used as input parameters, the effective mixing angle $\sin^2\theta_W^\textit{eff}$ is fixed at tree level as $(\sin^2\theta_W^\textit{eff}(1-\sin^2\theta_W^\textit{eff}))^{-1}=\pi\alpha_\text{em}/(\sqrt{2}G_Fm_Z^2)$. At one loop, it receives corrections proportional to $m_t^2$. Scenarios that generate electroweak symmetry breaking beyond the SM explanation do leave an imprint on $\sin^2\theta_W^\textit{eff}$ as well. A generic way to describe such effects is through the $\widehat{S}$ parameter, generated by new physics effects in the vacuum polarization of $SU(2)_L$ and $U(1)_Y$. This is also generated via the single dimension six term 
$\mathcal{L}\supset \widehat{S}\frac{g g^\prime}{4 m_W^2}\,H^\dagger \tau^a \overleftrightarrow{D}_\mu H W^a_{\mu\nu}B_{\mu\nu}$  \cite{Barbieri:2004qk,LHCHiggsCrossSectionWorkingGroup:2016ypw}. The parameter $\widehat{S}$ defines a microscopic scale $\Lambda^2\equiv m_W^2\times \widehat{S}^{-1}$, with the interpretation of the typical scale at which the effective interaction is generated.
We assume that this is the most relevant effect of new physics in order to evaluate the sensitivity. In particular, we neglect any lepton-flavor dependent effect, four-fermion interactions or other electroweak parameters. Therefore, under these assumptions deviations from the SM prediction of $\sin^2\theta_W^\textit{eff}$ are interpreted in solely terms of $\widehat{S}$, with contributions from $\delta\alpha_\text{em}$ and $\delta m_t$ as
\be
\frac{\delta \sin^2\theta_W^\textit{eff}}{\sin^2\theta_W^\textit{eff}}/10^{-5}\,\simeq\, -\frac{\delta(\alpha_\text{em}^{-1})}{10^{-3}}
-\frac{\delta m_t}{65\,\text{MeV}}
+\frac{\widehat{S}}{5\cdot 10^{-6}}\,.
\ee
The $A_{FB}^{\mu\mu}$ observable provides a $10^{-5}$ measurement of the effective mixing angle. 
Fully using such precision to constrain $\widehat{S}$ requires $\alpha_\text{em}$ and $m_t$ to be known at a level given by the expression. 
The ratios $\mathcal{R}_{e^-/\ell^\pm}$ provide sufficient sensitivity on $\alpha_\text{em}$, ensuring that it is no longer a bottleneck in interpreting the $\sin^2\theta_W^\textit{eff}$ measurement in terms of $\widehat{S}$. This is clear in Fig.~\ref{fig:aem_Sh}, where the measurements of $A_{FB}^{\mu\mu}$ and $\mathcal{R}_{e^-/\ell^\pm}$ are used to draw the sensitivity on the electromagnetic coupling and $\widehat{S}$. 
While the measurement of $\sin^2\theta_W^\textit{eff}$ from $A_{FB}^{\mu\mu}$ leads to a flat direction in the $\widehat{S}$ - $\alpha_\text{em}$ plane, adding $\mathcal{R}_{e^-/\ell^\pm}$ allows to constrain both parameters independently. 
Assuming the $17\,\text{MeV}$ reach on the top quark mass as obtained from the $t\bar{t}$ threshold run, one gets a sensitivity to scales up to $\Lambda\sim 40\,\text{TeV}$ from the combination of $A_{FB}^{\mu\mu}$ and $\mathcal{R}_{e^-/\ell^\pm}$.  We show for comparison the constraints using a sensitivity on $m_t$ expected from the HL-LHC. Due to the degeneracy between $m_t$ and $\widehat{S}$ when considering only $A_{FB}^{\mu\mu}$ and $\mathcal{R}_{e^-/\ell^\pm}$, the new physics reach is notably worse. The dashed lines correspond to $\delta m_t = 380\,\text{MeV}$, and the sensitivity to $\widehat{S}$ is barely above 10\,TeV. The scale reached raises as $\sqrt{\delta m_t}$, and lowering the top mass uncertainty to the 100\,MeV level increases the sensitivity up to $20\,\text{TeV}$.

\section{Conclusions}

Current measurements and future projections of an indirect determination of $\alpha_\text{em}(m_Z^2)$ are insufficient for the ambitious electroweak program of the Tera-$Z$ run at FCC-ee. In this work we propose observables at the $Z$-pole, $\mathcal{R}_{e^-/\mu^-}$ and $\mathcal{R}_{e^-/e^+}$, that have a relative statistical sensitivity to $\alpha_\text{em}(m_Z^2)$ below the $10^{-5}$ level, significantly improving over other methods. 

The ratios $\mathcal{R}_{e^-/\ell^\pm}$ offer robust sensitivity to $\alpha_\text{em}(m_Z^2)$, with the main sources of parametric uncertainty due to $\Delta\alpha(t,m_Z^2)$ and $m_t$ under control. 
If the statistical sensitivity of $\mathcal{R}_{e^-/\ell^\pm}$ on-peak is achieved, the bottleneck for interpreting measurements in terms of beyond SM effects is no longer $\alpha_\text{em}(m_Z^2)$. 
The $t\bar{t}$ threshold run at FCC-ee is crucial to bring the top mass uncertainty to a level that does not impact interpretations of Tera-$Z$ data. 
It would be interesting to embed the proposed observables into an electroweak global fit \cite{deBlas:2016nqo,deBlas:2022ofj}, 
explore the new physics reach in other input schemes, 
and to study the scenario of having Tera-$Z$ data without the $t\bar{t}$ run in order to set a goal for the top mass measurement at the HL-LHC.

We leave for future work the endeavor of studying $\mathcal{R}_{e^-/\ell^\pm}$ at higher orders in perturbation theory, assessing the extent of the cancellation of higher order effects, refining the treatment of hadronic vacuum polarization, and identifying the requirements to reduce theoretical uncertainties below the statistical sensitivity. 

The statistical precision of the Tera-$Z$ run is absolutely unprecedented in collider environments and represents a qualitative and transformative leap forward. The proposed observables are a step towards unlocking the potential of this precision.

\subsection*{Acknowledgments}

I am very grateful to Alain Blondel, Tim Cohen, Jorge de Blas, Gauthier Durieux, Majid Ekhterachian, Patrick Janot, Matthew McCullough, Aditya Pathak, Fulvio Piccinini, Francesco P. Ucci, Riccardo Rattazzi, Francesco Riva and Sebastian Wuchterl for useful discussions, suggestions and comments on the draft.

\bibliography{refsTZ} 

\begin{thebibliography}{84}%
\makeatletter
\providecommand \@ifxundefined [1]{%
 \@ifx{#1\undefined}
}%
\providecommand \@ifnum [1]{%
 \ifnum #1\expandafter \@firstoftwo
 \else \expandafter \@secondoftwo
 \fi
}%
\providecommand \@ifx [1]{%
 \ifx #1\expandafter \@firstoftwo
 \else \expandafter \@secondoftwo
 \fi
}%
\providecommand \natexlab [1]{#1}%
\providecommand \enquote  [1]{``#1''}%
\providecommand \bibnamefont  [1]{#1}%
\providecommand \bibfnamefont [1]{#1}%
\providecommand \citenamefont [1]{#1}%
\providecommand \href@noop [0]{\@secondoftwo}%
\providecommand \href [0]{\begingroup \@sanitize@url \@href}%
\providecommand \@href[1]{\@@startlink{#1}\@@href}%
\providecommand \@@href[1]{\endgroup#1\@@endlink}%
\providecommand \@sanitize@url [0]{\catcode `\\12\catcode `\$12\catcode
  `\&12\catcode `\#12\catcode `\^12\catcode `\_12\catcode `\%12\relax}%
\providecommand \@@startlink[1]{}%
\providecommand \@@endlink[0]{}%
\providecommand \url  [0]{\begingroup\@sanitize@url \@url }%
\providecommand \@url [1]{\endgroup\@href {#1}{\urlprefix }}%
\providecommand \urlprefix  [0]{URL }%
\providecommand \Eprint [0]{\href }%
\providecommand \doibase [0]{http://dx.doi.org/}%
\providecommand \selectlanguage [0]{\@gobble}%
\providecommand \bibinfo  [0]{\@secondoftwo}%
\providecommand \bibfield  [0]{\@secondoftwo}%
\providecommand \translation [1]{[#1]}%
\providecommand \BibitemOpen [0]{}%
\providecommand \bibitemStop [0]{}%
\providecommand \bibitemNoStop [0]{.\EOS\space}%
\providecommand \EOS [0]{\spacefactor3000\relax}%
\providecommand \BibitemShut  [1]{\csname bibitem#1\endcsname}%
\let\auto@bib@innerbib\@empty
\bibitem [{\citenamefont {Abada}\ \emph {et~al.}(2019)\citenamefont {Abada}
  \emph {et~al.}}]{FCC:2018evy}%
  \BibitemOpen
  \bibfield  {author} {\bibinfo {author} {\bibfnamefont {A.}~\bibnamefont
  {Abada}} \emph {et~al.} (\bibinfo {collaboration} {FCC}),\ }\href {\doibase
  10.1140/epjst/e2019-900045-4} {\bibfield  {journal} {\bibinfo  {journal}
  {Eur. Phys. J. ST}\ }\textbf {\bibinfo {volume} {228}},\ \bibinfo {pages}
  {261} (\bibinfo {year} {2019})}\BibitemShut {NoStop}%
\bibitem [{\citenamefont {Aoyama}\ \emph {et~al.}(2019)\citenamefont {Aoyama},
  \citenamefont {Kinoshita},\ and\ \citenamefont {Nio}}]{atoms7010028}%
  \BibitemOpen
  \bibfield  {author} {\bibinfo {author} {\bibfnamefont {T.}~\bibnamefont
  {Aoyama}}, \bibinfo {author} {\bibfnamefont {T.}~\bibnamefont {Kinoshita}}, \
  and\ \bibinfo {author} {\bibfnamefont {M.}~\bibnamefont {Nio}},\ }\href
  {\doibase 10.3390/atoms7010028} {\bibfield  {journal} {\bibinfo  {journal}
  {Atoms}\ }\textbf {\bibinfo {volume} {7}} (\bibinfo {year} {2019}),\
  10.3390/atoms7010028}\BibitemShut {NoStop}%
\bibitem [{\citenamefont {Steinhauser}(1998)}]{Steinhauser:1998rq}%
  \BibitemOpen
  \bibfield  {author} {\bibinfo {author} {\bibfnamefont {M.}~\bibnamefont
  {Steinhauser}},\ }\href {\doibase 10.1016/S0370-2693(98)00503-6} {\bibfield
  {journal} {\bibinfo  {journal} {Phys. Lett. B}\ }\textbf {\bibinfo {volume}
  {429}},\ \bibinfo {pages} {158} (\bibinfo {year} {1998})},\ \Eprint
  {http://arxiv.org/abs/hep-ph/9803313} {arXiv:hep-ph/9803313} \BibitemShut
  {NoStop}%
\bibitem [{\citenamefont {Sturm}(2013)}]{Sturm:2013uka}%
  \BibitemOpen
  \bibfield  {author} {\bibinfo {author} {\bibfnamefont {C.}~\bibnamefont
  {Sturm}},\ }\href {\doibase 10.1016/j.nuclphysb.2013.06.009} {\bibfield
  {journal} {\bibinfo  {journal} {Nucl. Phys. B}\ }\textbf {\bibinfo {volume}
  {874}},\ \bibinfo {pages} {698} (\bibinfo {year} {2013})},\ \Eprint
  {http://arxiv.org/abs/1305.0581} {arXiv:1305.0581 [hep-ph]} \BibitemShut
  {NoStop}%
\bibitem [{\citenamefont {Erler}\ and\ \citenamefont
  {Ferro-Hern\'andez}(2018)}]{Erler:2017knj}%
  \BibitemOpen
  \bibfield  {author} {\bibinfo {author} {\bibfnamefont {J.}~\bibnamefont
  {Erler}}\ and\ \bibinfo {author} {\bibfnamefont {R.}~\bibnamefont
  {Ferro-Hern\'andez}},\ }\href {\doibase 10.1007/JHEP03(2018)196} {\bibfield
  {journal} {\bibinfo  {journal} {JHEP}\ }\textbf {\bibinfo {volume} {03}},\
  \bibinfo {pages} {196} (\bibinfo {year} {2018})},\ \Eprint
  {http://arxiv.org/abs/1712.09146} {arXiv:1712.09146 [hep-ph]} \BibitemShut
  {NoStop}%
\bibitem [{\citenamefont {Jegerlehner}(2020)}]{Jegerlehner:2019lxt}%
  \BibitemOpen
  \bibfield  {author} {\bibinfo {author} {\bibfnamefont {F.}~\bibnamefont
  {Jegerlehner}},\ }\href {\doibase 10.23731/CYRM-2020-003.9} {\bibfield
  {journal} {\bibinfo  {journal} {CERN Yellow Reports: Monographs}\ }\textbf
  {\bibinfo {volume} {3}},\ \bibinfo {pages} {9} (\bibinfo {year}
  {2020})}\BibitemShut {NoStop}%
\bibitem [{\citenamefont {Blondel}\ \emph
  {et~al.}(2019{\natexlab{a}})\citenamefont {Blondel}, \citenamefont {Gluza},
  \citenamefont {Jadach}, \citenamefont {Janot},\ and\ \citenamefont
  {Riemann}}]{Proceedings:2019vxr}%
  \BibitemOpen
  \bibinfo {editor} {\bibfnamefont {A.}~\bibnamefont {Blondel}}, \bibinfo
  {editor} {\bibfnamefont {J.}~\bibnamefont {Gluza}}, \bibinfo {editor}
  {\bibfnamefont {S.}~\bibnamefont {Jadach}}, \bibinfo {editor} {\bibfnamefont
  {P.}~\bibnamefont {Janot}}, \ and\ \bibinfo {editor} {\bibfnamefont
  {T.}~\bibnamefont {Riemann}},\ eds.,\ \href {\doibase 10.23731/CYRM-2020-003}
  {\emph {\bibinfo {title} {{Theory for the FCC-ee}: {Report on the 11th FCC-ee
  Workshop Theory and Experiments}}}},\ \bibinfo {series} {CERN Yellow Reports:
  Monographs}, Vol.\ \bibinfo {volume} {3/2020}\ (\bibinfo  {publisher}
  {CERN},\ \bibinfo {address} {Geneva},\ \bibinfo {year} {2019})\ \Eprint
  {http://arxiv.org/abs/1905.05078} {arXiv:1905.05078 [hep-ph]} \BibitemShut
  {NoStop}%
\bibitem [{\citenamefont {Davier}\ \emph {et~al.}(2020)\citenamefont {Davier},
  \citenamefont {Hoecker}, \citenamefont {Malaescu},\ and\ \citenamefont
  {Zhang}}]{Davier:2019can}%
  \BibitemOpen
  \bibfield  {author} {\bibinfo {author} {\bibfnamefont {M.}~\bibnamefont
  {Davier}}, \bibinfo {author} {\bibfnamefont {A.}~\bibnamefont {Hoecker}},
  \bibinfo {author} {\bibfnamefont {B.}~\bibnamefont {Malaescu}}, \ and\
  \bibinfo {author} {\bibfnamefont {Z.}~\bibnamefont {Zhang}},\ }\href
  {\doibase 10.1140/epjc/s10052-020-7792-2} {\bibfield  {journal} {\bibinfo
  {journal} {Eur. Phys. J. C}\ }\textbf {\bibinfo {volume} {80}},\ \bibinfo
  {pages} {241} (\bibinfo {year} {2020})},\ \bibinfo {note} {[Erratum:
  Eur.Phys.J.C 80, 410 (2020)]},\ \Eprint {http://arxiv.org/abs/1908.00921}
  {arXiv:1908.00921 [hep-ph]} \BibitemShut {NoStop}%
\bibitem [{\citenamefont {Keshavarzi}\ \emph {et~al.}(2020)\citenamefont
  {Keshavarzi}, \citenamefont {Nomura},\ and\ \citenamefont
  {Teubner}}]{Keshavarzi:2019abf}%
  \BibitemOpen
  \bibfield  {author} {\bibinfo {author} {\bibfnamefont {A.}~\bibnamefont
  {Keshavarzi}}, \bibinfo {author} {\bibfnamefont {D.}~\bibnamefont {Nomura}},
  \ and\ \bibinfo {author} {\bibfnamefont {T.}~\bibnamefont {Teubner}},\ }\href
  {\doibase 10.1103/PhysRevD.101.014029} {\bibfield  {journal} {\bibinfo
  {journal} {Phys. Rev. D}\ }\textbf {\bibinfo {volume} {101}},\ \bibinfo
  {pages} {014029} (\bibinfo {year} {2020})},\ \Eprint
  {http://arxiv.org/abs/1911.00367} {arXiv:1911.00367 [hep-ph]} \BibitemShut
  {NoStop}%
\bibitem [{\citenamefont {C\`e}\ \emph {et~al.}(2022)\citenamefont {C\`e},
  \citenamefont {G\'erardin}, \citenamefont {von Hippel}, \citenamefont
  {Meyer}, \citenamefont {Miura}, \citenamefont {Ottnad}, \citenamefont
  {Risch}, \citenamefont {San~Jos\'e}, \citenamefont {Wilhelm},\ and\
  \citenamefont {Wittig}}]{Ce:2022eix}%
  \BibitemOpen
  \bibfield  {author} {\bibinfo {author} {\bibfnamefont {M.}~\bibnamefont
  {C\`e}}, \bibinfo {author} {\bibfnamefont {A.}~\bibnamefont {G\'erardin}},
  \bibinfo {author} {\bibfnamefont {G.}~\bibnamefont {von Hippel}}, \bibinfo
  {author} {\bibfnamefont {H.~B.}\ \bibnamefont {Meyer}}, \bibinfo {author}
  {\bibfnamefont {K.}~\bibnamefont {Miura}}, \bibinfo {author} {\bibfnamefont
  {K.}~\bibnamefont {Ottnad}}, \bibinfo {author} {\bibfnamefont
  {A.}~\bibnamefont {Risch}}, \bibinfo {author} {\bibfnamefont
  {T.}~\bibnamefont {San~Jos\'e}}, \bibinfo {author} {\bibfnamefont
  {J.}~\bibnamefont {Wilhelm}}, \ and\ \bibinfo {author} {\bibfnamefont
  {H.}~\bibnamefont {Wittig}},\ }\href {\doibase 10.1007/JHEP08(2022)220}
  {\bibfield  {journal} {\bibinfo  {journal} {JHEP}\ }\textbf {\bibinfo
  {volume} {08}},\ \bibinfo {pages} {220} (\bibinfo {year} {2022})},\ \Eprint
  {http://arxiv.org/abs/2203.08676} {arXiv:2203.08676 [hep-lat]} \BibitemShut
  {NoStop}%
\bibitem [{\citenamefont {Navas}\ \emph {et~al.}(2024)\citenamefont {Navas}
  \emph {et~al.}}]{ParticleDataGroup:2024cfk}%
  \BibitemOpen
  \bibfield  {author} {\bibinfo {author} {\bibfnamefont {S.}~\bibnamefont
  {Navas}} \emph {et~al.} (\bibinfo {collaboration} {Particle Data Group}),\
  }\href {\doibase 10.1103/PhysRevD.110.030001} {\bibfield  {journal} {\bibinfo
   {journal} {Phys. Rev. D}\ }\textbf {\bibinfo {volume} {110}},\ \bibinfo
  {pages} {030001} (\bibinfo {year} {2024})}\BibitemShut {NoStop}%
\bibitem [{\citenamefont {Davier}\ \emph {et~al.}(2024)\citenamefont {Davier},
  \citenamefont {Fodor}, \citenamefont {Gerardin}, \citenamefont {Lellouch},
  \citenamefont {Malaescu}, \citenamefont {Stokes}, \citenamefont {Szabo},
  \citenamefont {Toth}, \citenamefont {Varnhorst},\ and\ \citenamefont
  {Zhang}}]{Davier:2023cyp}%
  \BibitemOpen
  \bibfield  {author} {\bibinfo {author} {\bibfnamefont {M.}~\bibnamefont
  {Davier}}, \bibinfo {author} {\bibfnamefont {Z.}~\bibnamefont {Fodor}},
  \bibinfo {author} {\bibfnamefont {A.}~\bibnamefont {Gerardin}}, \bibinfo
  {author} {\bibfnamefont {L.}~\bibnamefont {Lellouch}}, \bibinfo {author}
  {\bibfnamefont {B.}~\bibnamefont {Malaescu}}, \bibinfo {author}
  {\bibfnamefont {F.~M.}\ \bibnamefont {Stokes}}, \bibinfo {author}
  {\bibfnamefont {K.~K.}\ \bibnamefont {Szabo}}, \bibinfo {author}
  {\bibfnamefont {B.~C.}\ \bibnamefont {Toth}}, \bibinfo {author}
  {\bibfnamefont {L.}~\bibnamefont {Varnhorst}}, \ and\ \bibinfo {author}
  {\bibfnamefont {Z.}~\bibnamefont {Zhang}},\ }\href {\doibase
  10.1103/PhysRevD.109.076019} {\bibfield  {journal} {\bibinfo  {journal}
  {Phys. Rev. D}\ }\textbf {\bibinfo {volume} {109}},\ \bibinfo {pages}
  {076019} (\bibinfo {year} {2024})},\ \Eprint
  {http://arxiv.org/abs/2308.04221} {arXiv:2308.04221 [hep-ph]} \BibitemShut
  {NoStop}%
\bibitem [{\citenamefont {Erler}\ \emph {et~al.}(2024)\citenamefont {Erler},
  \citenamefont {Ferro-Hernandez},\ and\ \citenamefont
  {Kuberski}}]{Erler:2024lds}%
  \BibitemOpen
  \bibfield  {author} {\bibinfo {author} {\bibfnamefont {J.}~\bibnamefont
  {Erler}}, \bibinfo {author} {\bibfnamefont {R.}~\bibnamefont
  {Ferro-Hernandez}}, \ and\ \bibinfo {author} {\bibfnamefont {S.}~\bibnamefont
  {Kuberski}},\ }\href {\doibase 10.1103/PhysRevLett.133.171801} {\bibfield
  {journal} {\bibinfo  {journal} {Phys. Rev. Lett.}\ }\textbf {\bibinfo
  {volume} {133}},\ \bibinfo {pages} {171801} (\bibinfo {year} {2024})},\
  \Eprint {http://arxiv.org/abs/2406.16691} {arXiv:2406.16691 [hep-ph]}
  \BibitemShut {NoStop}%
\bibitem [{\citenamefont {Janot}(2016)}]{Janot:2015gjr}%
  \BibitemOpen
  \bibfield  {author} {\bibinfo {author} {\bibfnamefont {P.}~\bibnamefont
  {Janot}},\ }\href {\doibase 10.1007/JHEP02(2016)053} {\bibfield  {journal}
  {\bibinfo  {journal} {JHEP}\ }\textbf {\bibinfo {volume} {02}},\ \bibinfo
  {pages} {053} (\bibinfo {year} {2016})},\ \bibinfo {note} {[Erratum: JHEP 11,
  164 (2017)]},\ \Eprint {http://arxiv.org/abs/1512.05544} {arXiv:1512.05544
  [hep-ph]} \BibitemShut {NoStop}%
\bibitem [{\citenamefont {Beenakker}\ \emph {et~al.}(1991)\citenamefont
  {Beenakker}, \citenamefont {Berends},\ and\ \citenamefont {van~der
  Marck}}]{Beenakker:1990mb}%
  \BibitemOpen
  \bibfield  {author} {\bibinfo {author} {\bibfnamefont {W.}~\bibnamefont
  {Beenakker}}, \bibinfo {author} {\bibfnamefont {F.~A.}\ \bibnamefont
  {Berends}}, \ and\ \bibinfo {author} {\bibfnamefont {S.~C.}\ \bibnamefont
  {van~der Marck}},\ }\href {\doibase 10.1016/0550-3213(91)90328-U} {\bibfield
  {journal} {\bibinfo  {journal} {Nucl. Phys. B}\ }\textbf {\bibinfo {volume}
  {349}},\ \bibinfo {pages} {323} (\bibinfo {year} {1991})}\BibitemShut
  {NoStop}%
\bibitem [{\citenamefont {Beenakker}\ and\ \citenamefont
  {Passarino}(1998)}]{BEENAKKER1998199}%
  \BibitemOpen
  \bibfield  {author} {\bibinfo {author} {\bibfnamefont {W.}~\bibnamefont
  {Beenakker}}\ and\ \bibinfo {author} {\bibfnamefont {G.}~\bibnamefont
  {Passarino}},\ }\href {\doibase
  https://doi.org/10.1016/S0370-2693(98)00176-2} {\bibfield  {journal}
  {\bibinfo  {journal} {Physics Letters B}\ }\textbf {\bibinfo {volume}
  {425}},\ \bibinfo {pages} {199} (\bibinfo {year} {1998})}\BibitemShut
  {NoStop}%
\bibitem [{\citenamefont {Placzek}\ \emph {et~al.}(1999)\citenamefont
  {Placzek}, \citenamefont {Jadach}, \citenamefont {Melles}, \citenamefont
  {Ward},\ and\ \citenamefont {Yost}}]{Placzek:1999xc}%
  \BibitemOpen
  \bibfield  {author} {\bibinfo {author} {\bibfnamefont {W.}~\bibnamefont
  {Placzek}}, \bibinfo {author} {\bibfnamefont {S.}~\bibnamefont {Jadach}},
  \bibinfo {author} {\bibfnamefont {M.}~\bibnamefont {Melles}}, \bibinfo
  {author} {\bibfnamefont {B.~F.~L.}\ \bibnamefont {Ward}}, \ and\ \bibinfo
  {author} {\bibfnamefont {S.~A.}\ \bibnamefont {Yost}},\ }in\ \href@noop {}
  {\emph {\bibinfo {booktitle} {{4th International Symposium on Radiative
  Corrections: Applications of Quantum Field Theory to Phenomenology}}}}\
  (\bibinfo {year} {1999})\ pp.\ \bibinfo {pages} {325--333},\ \Eprint
  {http://arxiv.org/abs/hep-ph/9903381} {arXiv:hep-ph/9903381} \BibitemShut
  {NoStop}%
\bibitem [{\citenamefont {Montagna}\ \emph {et~al.}(1999)\citenamefont
  {Montagna}, \citenamefont {Nicrosini},\ and\ \citenamefont
  {Piccinini}}]{Montagna:1999tf}%
  \BibitemOpen
  \bibfield  {author} {\bibinfo {author} {\bibfnamefont {G.}~\bibnamefont
  {Montagna}}, \bibinfo {author} {\bibfnamefont {O.}~\bibnamefont {Nicrosini}},
  \ and\ \bibinfo {author} {\bibfnamefont {F.}~\bibnamefont {Piccinini}},\
  }\href {\doibase 10.1016/S0370-2693(99)00791-1} {\bibfield  {journal}
  {\bibinfo  {journal} {Phys. Lett. B}\ }\textbf {\bibinfo {volume} {460}},\
  \bibinfo {pages} {425} (\bibinfo {year} {1999})},\ \Eprint
  {http://arxiv.org/abs/hep-ph/9904387} {arXiv:hep-ph/9904387} \BibitemShut
  {NoStop}%
\bibitem [{\citenamefont {Jadach}\ \emph {et~al.}(1995)\citenamefont {Jadach},
  \citenamefont {Płaczek},\ and\ \citenamefont {Ward}}]{JADACH1995349}%
  \BibitemOpen
  \bibfield  {author} {\bibinfo {author} {\bibfnamefont {S.}~\bibnamefont
  {Jadach}}, \bibinfo {author} {\bibfnamefont {W.}~\bibnamefont {Płaczek}}, \
  and\ \bibinfo {author} {\bibfnamefont {B.}~\bibnamefont {Ward}},\ }\href
  {\doibase https://doi.org/10.1016/0370-2693(95)00576-7} {\bibfield  {journal}
  {\bibinfo  {journal} {Physics Letters B}\ }\textbf {\bibinfo {volume}
  {353}},\ \bibinfo {pages} {349} (\bibinfo {year} {1995})}\BibitemShut
  {NoStop}%
\bibitem [{\citenamefont {Carloni~Calame}\ \emph {et~al.}(2015)\citenamefont
  {Carloni~Calame}, \citenamefont {Montagna}, \citenamefont {Nicrosini},\ and\
  \citenamefont {Piccinini}}]{CarloniCalame:2015zev}%
  \BibitemOpen
  \bibfield  {author} {\bibinfo {author} {\bibfnamefont {C.~M.}\ \bibnamefont
  {Carloni~Calame}}, \bibinfo {author} {\bibfnamefont {G.}~\bibnamefont
  {Montagna}}, \bibinfo {author} {\bibfnamefont {O.}~\bibnamefont {Nicrosini}},
  \ and\ \bibinfo {author} {\bibfnamefont {F.}~\bibnamefont {Piccinini}},\
  }\href {\doibase 10.5506/APhysPolB.46.2227} {\bibfield  {journal} {\bibinfo
  {journal} {Acta Phys. Polon. B}\ }\textbf {\bibinfo {volume} {46}},\ \bibinfo
  {pages} {2227} (\bibinfo {year} {2015})}\BibitemShut {NoStop}%
\bibitem [{\citenamefont {Jadach}\ \emph {et~al.}(2019)\citenamefont {Jadach},
  \citenamefont {P\l{}aczek}, \citenamefont {Skrzypek}, \citenamefont {Ward},\
  and\ \citenamefont {Yost}}]{Jadach:2018jjo}%
  \BibitemOpen
  \bibfield  {author} {\bibinfo {author} {\bibfnamefont {S.}~\bibnamefont
  {Jadach}}, \bibinfo {author} {\bibfnamefont {W.}~\bibnamefont {P\l{}aczek}},
  \bibinfo {author} {\bibfnamefont {M.}~\bibnamefont {Skrzypek}}, \bibinfo
  {author} {\bibfnamefont {B.~F.~L.}\ \bibnamefont {Ward}}, \ and\ \bibinfo
  {author} {\bibfnamefont {S.~A.}\ \bibnamefont {Yost}},\ }\href {\doibase
  10.1016/j.physletb.2019.01.012} {\bibfield  {journal} {\bibinfo  {journal}
  {Phys. Lett. B}\ }\textbf {\bibinfo {volume} {790}},\ \bibinfo {pages} {314}
  (\bibinfo {year} {2019})},\ \Eprint {http://arxiv.org/abs/1812.01004}
  {arXiv:1812.01004 [hep-ph]} \BibitemShut {NoStop}%
\bibitem [{\citenamefont {Dam}(2022)}]{Dam:2021sdj}%
  \BibitemOpen
  \bibfield  {author} {\bibinfo {author} {\bibfnamefont {M.}~\bibnamefont
  {Dam}},\ }\href {\doibase 10.1140/epjp/s13360-021-02265-3} {\bibfield
  {journal} {\bibinfo  {journal} {Eur. Phys. J. Plus}\ }\textbf {\bibinfo
  {volume} {137}},\ \bibinfo {pages} {81} (\bibinfo {year} {2022})},\ \Eprint
  {http://arxiv.org/abs/2107.12837} {arXiv:2107.12837 [physics.ins-det]}
  \BibitemShut {NoStop}%
\bibitem [{\citenamefont {Consoli}(1979)}]{CONSOLI1979208}%
  \BibitemOpen
  \bibfield  {author} {\bibinfo {author} {\bibfnamefont {M.}~\bibnamefont
  {Consoli}},\ }\href {\doibase https://doi.org/10.1016/0550-3213(79)90235-9}
  {\bibfield  {journal} {\bibinfo  {journal} {Nuclear Physics B}\ }\textbf
  {\bibinfo {volume} {160}},\ \bibinfo {pages} {208} (\bibinfo {year}
  {1979})}\BibitemShut {NoStop}%
\bibitem [{\citenamefont {Caffo}\ and\ \citenamefont
  {Remiddi}(1989)}]{Caffo:367852}%
  \BibitemOpen
  \bibfield  {author} {\bibinfo {author} {\bibfnamefont {M.}~\bibnamefont
  {Caffo}}\ and\ \bibinfo {author} {\bibfnamefont {E.}~\bibnamefont
  {Remiddi}},\ }\href {\doibase 10.5170/CERN-1989-008-V-1.171} {\  (\bibinfo
  {year} {1989}),\ 10.5170/CERN-1989-008-V-1.171}\BibitemShut {NoStop}%
\bibitem [{\citenamefont {Altarelli}\ \emph {et~al.}(1989)\citenamefont
  {Altarelli}, \citenamefont {Kleiss},\ and\ \citenamefont
  {Verzegnassi}}]{Altarelli:1989hv}%
  \BibitemOpen
  \bibinfo {editor} {\bibfnamefont {G.}~\bibnamefont {Altarelli}}, \bibinfo
  {editor} {\bibfnamefont {R.}~\bibnamefont {Kleiss}}, \ and\ \bibinfo {editor}
  {\bibfnamefont {C.}~\bibnamefont {Verzegnassi}},\ eds.,\ \href {\doibase
  10.5170/CERN-1989-008-V-1} {\emph {\bibinfo {title} {{Z PHYSICS AT LEP-1.
  PROCEEDINGS, WORKSHOP, GENEVA, SWITZERLAND, SEPTEMBER 4-5, 1989. VOL. 1:
  STANDARD PHYSICS}}}},\ CERN Yellow Reports: Conference Proceedings\ (\bibinfo
  {year} {1989})\BibitemShut {NoStop}%
\bibitem [{\citenamefont {Barchetta}\ \emph {et~al.}(2022)\citenamefont
  {Barchetta}, \citenamefont {Collins},\ and\ \citenamefont
  {Riedler}}]{Barchetta:2021ibt}%
  \BibitemOpen
  \bibfield  {author} {\bibinfo {author} {\bibfnamefont {N.}~\bibnamefont
  {Barchetta}}, \bibinfo {author} {\bibfnamefont {P.}~\bibnamefont {Collins}},
  \ and\ \bibinfo {author} {\bibfnamefont {P.}~\bibnamefont {Riedler}},\ }\href
  {\doibase 10.1140/epjp/s13360-021-02323-w} {\bibfield  {journal} {\bibinfo
  {journal} {Eur. Phys. J. Plus}\ }\textbf {\bibinfo {volume} {137}},\ \bibinfo
  {pages} {231} (\bibinfo {year} {2022})},\ \Eprint
  {http://arxiv.org/abs/2112.13019} {arXiv:2112.13019 [physics.ins-det]}
  \BibitemShut {NoStop}%
\bibitem [{\citenamefont {Berends}\ and\ \citenamefont
  {Kleiss}(1983)}]{BERENDS1983537}%
  \BibitemOpen
  \bibfield  {author} {\bibinfo {author} {\bibfnamefont {F.}~\bibnamefont
  {Berends}}\ and\ \bibinfo {author} {\bibfnamefont {R.}~\bibnamefont
  {Kleiss}},\ }\href {\doibase https://doi.org/10.1016/0550-3213(83)90558-8}
  {\bibfield  {journal} {\bibinfo  {journal} {Nuclear Physics B}\ }\textbf
  {\bibinfo {volume} {228}},\ \bibinfo {pages} {537} (\bibinfo {year}
  {1983})}\BibitemShut {NoStop}%
\bibitem [{\citenamefont {Caffo}\ \emph {et~al.}(1985)\citenamefont {Caffo},
  \citenamefont {Gatto},\ and\ \citenamefont {Remiddi}}]{CAFFO1985378}%
  \BibitemOpen
  \bibfield  {author} {\bibinfo {author} {\bibfnamefont {M.}~\bibnamefont
  {Caffo}}, \bibinfo {author} {\bibfnamefont {R.}~\bibnamefont {Gatto}}, \ and\
  \bibinfo {author} {\bibfnamefont {E.}~\bibnamefont {Remiddi}},\ }\href
  {\doibase https://doi.org/10.1016/0550-3213(85)90453-5} {\bibfield  {journal}
  {\bibinfo  {journal} {Nuclear Physics B}\ }\textbf {\bibinfo {volume}
  {252}},\ \bibinfo {pages} {378} (\bibinfo {year} {1985})}\BibitemShut
  {NoStop}%
\bibitem [{\citenamefont {Tobimatsu}\ and\ \citenamefont
  {Shimizu}(1985)}]{Tobimatsu:1985vd}%
  \BibitemOpen
  \bibfield  {author} {\bibinfo {author} {\bibfnamefont {K.}~\bibnamefont
  {Tobimatsu}}\ and\ \bibinfo {author} {\bibfnamefont {Y.}~\bibnamefont
  {Shimizu}},\ }\href {\doibase 10.1143/PTP.74.567} {\bibfield  {journal}
  {\bibinfo  {journal} {Prog. Theor. Phys.}\ }\textbf {\bibinfo {volume}
  {74}},\ \bibinfo {pages} {567} (\bibinfo {year} {1985})},\ \bibinfo {note}
  {[Erratum: Prog.Theor.Phys. 76, 334 (1986)]}\BibitemShut {NoStop}%
\bibitem [{\citenamefont {Tobimatsu}\ and\ \citenamefont
  {Shimizu}(1986)}]{Tobimatsu:1985pp}%
  \BibitemOpen
  \bibfield  {author} {\bibinfo {author} {\bibfnamefont {K.}~\bibnamefont
  {Tobimatsu}}\ and\ \bibinfo {author} {\bibfnamefont {Y.}~\bibnamefont
  {Shimizu}},\ }\href {\doibase 10.1143/PTP.75.905} {\bibfield  {journal}
  {\bibinfo  {journal} {Prog. Theor. Phys.}\ }\textbf {\bibinfo {volume}
  {75}},\ \bibinfo {pages} {905} (\bibinfo {year} {1986})}\BibitemShut
  {NoStop}%
\bibitem [{\citenamefont {Böhm}\ \emph {et~al.}(1988)\citenamefont {Böhm},
  \citenamefont {Denner},\ and\ \citenamefont {Hollik}}]{BOHM1988687}%
  \BibitemOpen
  \bibfield  {author} {\bibinfo {author} {\bibfnamefont {M.}~\bibnamefont
  {Böhm}}, \bibinfo {author} {\bibfnamefont {A.}~\bibnamefont {Denner}}, \
  and\ \bibinfo {author} {\bibfnamefont {W.}~\bibnamefont {Hollik}},\ }\href
  {\doibase https://doi.org/10.1016/0550-3213(88)90650-5} {\bibfield  {journal}
  {\bibinfo  {journal} {Nuclear Physics B}\ }\textbf {\bibinfo {volume}
  {304}},\ \bibinfo {pages} {687} (\bibinfo {year} {1988})}\BibitemShut
  {NoStop}%
\bibitem [{\citenamefont {Kuhn}\ \emph {et~al.}(2001)\citenamefont {Kuhn},
  \citenamefont {Moch}, \citenamefont {Penin},\ and\ \citenamefont
  {Smirnov}}]{Kuhn:2001hz}%
  \BibitemOpen
  \bibfield  {author} {\bibinfo {author} {\bibfnamefont {J.~H.}\ \bibnamefont
  {Kuhn}}, \bibinfo {author} {\bibfnamefont {S.}~\bibnamefont {Moch}}, \bibinfo
  {author} {\bibfnamefont {A.~A.}\ \bibnamefont {Penin}}, \ and\ \bibinfo
  {author} {\bibfnamefont {V.~A.}\ \bibnamefont {Smirnov}},\ }\href {\doibase
  10.1016/S0550-3213(01)00454-0} {\bibfield  {journal} {\bibinfo  {journal}
  {Nucl. Phys. B}\ }\textbf {\bibinfo {volume} {616}},\ \bibinfo {pages} {286}
  (\bibinfo {year} {2001})},\ \bibinfo {note} {[Erratum: Nucl.Phys.B 648,
  455--456 (2003)]},\ \Eprint {http://arxiv.org/abs/hep-ph/0106298}
  {arXiv:hep-ph/0106298} \BibitemShut {NoStop}%
\bibitem [{\citenamefont {Feucht}\ \emph {et~al.}(2004)\citenamefont {Feucht},
  \citenamefont {Kuhn}, \citenamefont {Penin},\ and\ \citenamefont
  {Smirnov}}]{Feucht:2004rp}%
  \BibitemOpen
  \bibfield  {author} {\bibinfo {author} {\bibfnamefont {B.}~\bibnamefont
  {Feucht}}, \bibinfo {author} {\bibfnamefont {J.~H.}\ \bibnamefont {Kuhn}},
  \bibinfo {author} {\bibfnamefont {A.~A.}\ \bibnamefont {Penin}}, \ and\
  \bibinfo {author} {\bibfnamefont {V.~A.}\ \bibnamefont {Smirnov}},\ }\href
  {\doibase 10.1103/PhysRevLett.93.101802} {\bibfield  {journal} {\bibinfo
  {journal} {Phys. Rev. Lett.}\ }\textbf {\bibinfo {volume} {93}},\ \bibinfo
  {pages} {101802} (\bibinfo {year} {2004})},\ \Eprint
  {http://arxiv.org/abs/hep-ph/0404082} {arXiv:hep-ph/0404082} \BibitemShut
  {NoStop}%
\bibitem [{\citenamefont {Jantzen}\ \emph {et~al.}(2005)\citenamefont
  {Jantzen}, \citenamefont {Kuhn}, \citenamefont {Penin},\ and\ \citenamefont
  {Smirnov}}]{Jantzen:2005az}%
  \BibitemOpen
  \bibfield  {author} {\bibinfo {author} {\bibfnamefont {B.}~\bibnamefont
  {Jantzen}}, \bibinfo {author} {\bibfnamefont {J.~H.}\ \bibnamefont {Kuhn}},
  \bibinfo {author} {\bibfnamefont {A.~A.}\ \bibnamefont {Penin}}, \ and\
  \bibinfo {author} {\bibfnamefont {V.~A.}\ \bibnamefont {Smirnov}},\ }\href
  {\doibase 10.1016/j.nuclphysb.2005.10.010} {\bibfield  {journal} {\bibinfo
  {journal} {Nucl. Phys. B}\ }\textbf {\bibinfo {volume} {731}},\ \bibinfo
  {pages} {188} (\bibinfo {year} {2005})},\ \bibinfo {note} {[Erratum:
  Nucl.Phys.B 752, 327--328 (2006)]},\ \Eprint
  {http://arxiv.org/abs/hep-ph/0509157} {arXiv:hep-ph/0509157} \BibitemShut
  {NoStop}%
\bibitem [{\citenamefont {Penin}\ and\ \citenamefont
  {Ryan}(2011)}]{Penin:2011aa}%
  \BibitemOpen
  \bibfield  {author} {\bibinfo {author} {\bibfnamefont {A.~A.}\ \bibnamefont
  {Penin}}\ and\ \bibinfo {author} {\bibfnamefont {G.}~\bibnamefont {Ryan}},\
  }\href {\doibase 10.1007/JHEP11(2011)081} {\bibfield  {journal} {\bibinfo
  {journal} {JHEP}\ }\textbf {\bibinfo {volume} {11}},\ \bibinfo {pages} {081}
  (\bibinfo {year} {2011})},\ \Eprint {http://arxiv.org/abs/1112.2171}
  {arXiv:1112.2171 [hep-ph]} \BibitemShut {NoStop}%
\bibitem [{\citenamefont {Bern}\ \emph {et~al.}(2001)\citenamefont {Bern},
  \citenamefont {Dixon},\ and\ \citenamefont {Ghinculov}}]{Bern:2000ie}%
  \BibitemOpen
  \bibfield  {author} {\bibinfo {author} {\bibfnamefont {Z.}~\bibnamefont
  {Bern}}, \bibinfo {author} {\bibfnamefont {L.~J.}\ \bibnamefont {Dixon}}, \
  and\ \bibinfo {author} {\bibfnamefont {A.}~\bibnamefont {Ghinculov}},\ }\href
  {\doibase 10.1103/PhysRevD.63.053007} {\bibfield  {journal} {\bibinfo
  {journal} {Phys. Rev. D}\ }\textbf {\bibinfo {volume} {63}},\ \bibinfo
  {pages} {053007} (\bibinfo {year} {2001})},\ \Eprint
  {http://arxiv.org/abs/hep-ph/0010075} {arXiv:hep-ph/0010075} \BibitemShut
  {NoStop}%
\bibitem [{\citenamefont {Penin}(2006)}]{Penin:2005eh}%
  \BibitemOpen
  \bibfield  {author} {\bibinfo {author} {\bibfnamefont {A.~A.}\ \bibnamefont
  {Penin}},\ }\href {\doibase 10.1016/j.nuclphysb.2005.11.016} {\bibfield
  {journal} {\bibinfo  {journal} {Nucl. Phys. B}\ }\textbf {\bibinfo {volume}
  {734}},\ \bibinfo {pages} {185} (\bibinfo {year} {2006})},\ \Eprint
  {http://arxiv.org/abs/hep-ph/0508127} {arXiv:hep-ph/0508127} \BibitemShut
  {NoStop}%
\bibitem [{\citenamefont {Mitov}\ and\ \citenamefont
  {Moch}(2007)}]{Mitov:2006xs}%
  \BibitemOpen
  \bibfield  {author} {\bibinfo {author} {\bibfnamefont {A.}~\bibnamefont
  {Mitov}}\ and\ \bibinfo {author} {\bibfnamefont {S.}~\bibnamefont {Moch}},\
  }\href {\doibase 10.1088/1126-6708/2007/05/001} {\bibfield  {journal}
  {\bibinfo  {journal} {JHEP}\ }\textbf {\bibinfo {volume} {05}},\ \bibinfo
  {pages} {001} (\bibinfo {year} {2007})},\ \Eprint
  {http://arxiv.org/abs/hep-ph/0612149} {arXiv:hep-ph/0612149} \BibitemShut
  {NoStop}%
\bibitem [{\citenamefont {Becher}\ and\ \citenamefont
  {Melnikov}(2007)}]{Becher:2007cu}%
  \BibitemOpen
  \bibfield  {author} {\bibinfo {author} {\bibfnamefont {T.}~\bibnamefont
  {Becher}}\ and\ \bibinfo {author} {\bibfnamefont {K.}~\bibnamefont
  {Melnikov}},\ }\href {\doibase 10.1088/1126-6708/2007/06/084} {\bibfield
  {journal} {\bibinfo  {journal} {JHEP}\ }\textbf {\bibinfo {volume} {06}},\
  \bibinfo {pages} {084} (\bibinfo {year} {2007})},\ \Eprint
  {http://arxiv.org/abs/0704.3582} {arXiv:0704.3582 [hep-ph]} \BibitemShut
  {NoStop}%
\bibitem [{\citenamefont {Actis}\ \emph {et~al.}(2007)\citenamefont {Actis},
  \citenamefont {Czakon}, \citenamefont {Gluza},\ and\ \citenamefont
  {Riemann}}]{Actis:2007gi}%
  \BibitemOpen
  \bibfield  {author} {\bibinfo {author} {\bibfnamefont {S.}~\bibnamefont
  {Actis}}, \bibinfo {author} {\bibfnamefont {M.}~\bibnamefont {Czakon}},
  \bibinfo {author} {\bibfnamefont {J.}~\bibnamefont {Gluza}}, \ and\ \bibinfo
  {author} {\bibfnamefont {T.}~\bibnamefont {Riemann}},\ }\href {\doibase
  10.1016/j.nuclphysb.2007.06.023} {\bibfield  {journal} {\bibinfo  {journal}
  {Nucl. Phys. B}\ }\textbf {\bibinfo {volume} {786}},\ \bibinfo {pages} {26}
  (\bibinfo {year} {2007})},\ \Eprint {http://arxiv.org/abs/0704.2400}
  {arXiv:0704.2400 [hep-ph]} \BibitemShut {NoStop}%
\bibitem [{\citenamefont {Bonciani}\ \emph {et~al.}(2004)\citenamefont
  {Bonciani}, \citenamefont {Ferroglia}, \citenamefont {Mastrolia},
  \citenamefont {Remiddi},\ and\ \citenamefont {van~der
  Bij}}]{Bonciani:2004gi}%
  \BibitemOpen
  \bibfield  {author} {\bibinfo {author} {\bibfnamefont {R.}~\bibnamefont
  {Bonciani}}, \bibinfo {author} {\bibfnamefont {A.}~\bibnamefont {Ferroglia}},
  \bibinfo {author} {\bibfnamefont {P.}~\bibnamefont {Mastrolia}}, \bibinfo
  {author} {\bibfnamefont {E.}~\bibnamefont {Remiddi}}, \ and\ \bibinfo
  {author} {\bibfnamefont {J.~J.}\ \bibnamefont {van~der Bij}},\ }\href
  {\doibase 10.1016/j.nuclphysb.2004.09.015} {\bibfield  {journal} {\bibinfo
  {journal} {Nucl. Phys. B}\ }\textbf {\bibinfo {volume} {701}},\ \bibinfo
  {pages} {121} (\bibinfo {year} {2004})},\ \Eprint
  {http://arxiv.org/abs/hep-ph/0405275} {arXiv:hep-ph/0405275} \BibitemShut
  {NoStop}%
\bibitem [{\citenamefont {Czakon}\ \emph {et~al.}(2006)\citenamefont {Czakon},
  \citenamefont {Gluza},\ and\ \citenamefont {Riemann}}]{Czakon:2006pa}%
  \BibitemOpen
  \bibfield  {author} {\bibinfo {author} {\bibfnamefont {M.}~\bibnamefont
  {Czakon}}, \bibinfo {author} {\bibfnamefont {J.}~\bibnamefont {Gluza}}, \
  and\ \bibinfo {author} {\bibfnamefont {T.}~\bibnamefont {Riemann}},\ }\href
  {\doibase 10.1016/j.nuclphysb.2006.05.033} {\bibfield  {journal} {\bibinfo
  {journal} {Nucl. Phys. B}\ }\textbf {\bibinfo {volume} {751}},\ \bibinfo
  {pages} {1} (\bibinfo {year} {2006})},\ \Eprint
  {http://arxiv.org/abs/hep-ph/0604101} {arXiv:hep-ph/0604101} \BibitemShut
  {NoStop}%
\bibitem [{\citenamefont {Actis}\ \emph {et~al.}(2010)\citenamefont {Actis},
  \citenamefont {Mastrolia},\ and\ \citenamefont {Ossola}}]{Actis:2009uq}%
  \BibitemOpen
  \bibfield  {author} {\bibinfo {author} {\bibfnamefont {S.}~\bibnamefont
  {Actis}}, \bibinfo {author} {\bibfnamefont {P.}~\bibnamefont {Mastrolia}}, \
  and\ \bibinfo {author} {\bibfnamefont {G.}~\bibnamefont {Ossola}},\ }\href
  {\doibase 10.1016/j.physletb.2009.11.035} {\bibfield  {journal} {\bibinfo
  {journal} {Phys. Lett. B}\ }\textbf {\bibinfo {volume} {682}},\ \bibinfo
  {pages} {419} (\bibinfo {year} {2010})},\ \Eprint
  {http://arxiv.org/abs/0909.1750} {arXiv:0909.1750 [hep-ph]} \BibitemShut
  {NoStop}%
\bibitem [{\citenamefont {Henn}\ and\ \citenamefont
  {Smirnov}(2013)}]{Henn:2013woa}%
  \BibitemOpen
  \bibfield  {author} {\bibinfo {author} {\bibfnamefont {J.~M.}\ \bibnamefont
  {Henn}}\ and\ \bibinfo {author} {\bibfnamefont {V.~A.}\ \bibnamefont
  {Smirnov}},\ }\href {\doibase 10.1007/JHEP11(2013)041} {\bibfield  {journal}
  {\bibinfo  {journal} {JHEP}\ }\textbf {\bibinfo {volume} {11}},\ \bibinfo
  {pages} {041} (\bibinfo {year} {2013})},\ \Eprint
  {http://arxiv.org/abs/1307.4083} {arXiv:1307.4083 [hep-th]} \BibitemShut
  {NoStop}%
\bibitem [{\citenamefont {Duhr}\ \emph {et~al.}(2021)\citenamefont {Duhr},
  \citenamefont {Smirnov},\ and\ \citenamefont {Tancredi}}]{Duhr:2021fhk}%
  \BibitemOpen
  \bibfield  {author} {\bibinfo {author} {\bibfnamefont {C.}~\bibnamefont
  {Duhr}}, \bibinfo {author} {\bibfnamefont {V.~A.}\ \bibnamefont {Smirnov}}, \
  and\ \bibinfo {author} {\bibfnamefont {L.}~\bibnamefont {Tancredi}},\ }\href
  {\doibase 10.1007/JHEP09(2021)120} {\bibfield  {journal} {\bibinfo  {journal}
  {JHEP}\ }\textbf {\bibinfo {volume} {09}},\ \bibinfo {pages} {120} (\bibinfo
  {year} {2021})},\ \Eprint {http://arxiv.org/abs/2108.03828} {arXiv:2108.03828
  [hep-ph]} \BibitemShut {NoStop}%
\bibitem [{\citenamefont {Delto}\ \emph {et~al.}(2024)\citenamefont {Delto},
  \citenamefont {Duhr}, \citenamefont {Tancredi},\ and\ \citenamefont
  {Zhu}}]{Delto:2023kqv}%
  \BibitemOpen
  \bibfield  {author} {\bibinfo {author} {\bibfnamefont {M.}~\bibnamefont
  {Delto}}, \bibinfo {author} {\bibfnamefont {C.}~\bibnamefont {Duhr}},
  \bibinfo {author} {\bibfnamefont {L.}~\bibnamefont {Tancredi}}, \ and\
  \bibinfo {author} {\bibfnamefont {Y.~J.}\ \bibnamefont {Zhu}},\ }\href
  {\doibase 10.1103/PhysRevLett.132.231904} {\bibfield  {journal} {\bibinfo
  {journal} {Phys. Rev. Lett.}\ }\textbf {\bibinfo {volume} {132}},\ \bibinfo
  {pages} {231904} (\bibinfo {year} {2024})},\ \Eprint
  {http://arxiv.org/abs/2311.06385} {arXiv:2311.06385 [hep-ph]} \BibitemShut
  {NoStop}%
\bibitem [{\citenamefont {Actis}\ \emph {et~al.}(2008)\citenamefont {Actis},
  \citenamefont {Czakon}, \citenamefont {Gluza},\ and\ \citenamefont
  {Riemann}}]{Actis:2007fs}%
  \BibitemOpen
  \bibfield  {author} {\bibinfo {author} {\bibfnamefont {S.}~\bibnamefont
  {Actis}}, \bibinfo {author} {\bibfnamefont {M.}~\bibnamefont {Czakon}},
  \bibinfo {author} {\bibfnamefont {J.}~\bibnamefont {Gluza}}, \ and\ \bibinfo
  {author} {\bibfnamefont {T.}~\bibnamefont {Riemann}},\ }\href {\doibase
  10.1103/PhysRevLett.100.131602} {\bibfield  {journal} {\bibinfo  {journal}
  {Phys. Rev. Lett.}\ }\textbf {\bibinfo {volume} {100}},\ \bibinfo {pages}
  {131602} (\bibinfo {year} {2008})},\ \Eprint {http://arxiv.org/abs/0711.3847}
  {arXiv:0711.3847 [hep-ph]} \BibitemShut {NoStop}%
\bibitem [{\citenamefont {Kuhn}\ and\ \citenamefont
  {Uccirati}(2009)}]{Kuhn:2008zs}%
  \BibitemOpen
  \bibfield  {author} {\bibinfo {author} {\bibfnamefont {J.~H.}\ \bibnamefont
  {Kuhn}}\ and\ \bibinfo {author} {\bibfnamefont {S.}~\bibnamefont
  {Uccirati}},\ }\href {\doibase 10.1016/j.nuclphysb.2008.08.002} {\bibfield
  {journal} {\bibinfo  {journal} {Nucl. Phys. B}\ }\textbf {\bibinfo {volume}
  {806}},\ \bibinfo {pages} {300} (\bibinfo {year} {2009})},\ \Eprint
  {http://arxiv.org/abs/0807.1284} {arXiv:0807.1284 [hep-ph]} \BibitemShut
  {NoStop}%
\bibitem [{\citenamefont {Sadykov}\ and\ \citenamefont
  {Yermolchyk}(2020)}]{Sadykov:2020any}%
  \BibitemOpen
  \bibfield  {author} {\bibinfo {author} {\bibfnamefont {R.}~\bibnamefont
  {Sadykov}}\ and\ \bibinfo {author} {\bibfnamefont {V.}~\bibnamefont
  {Yermolchyk}},\ }\href {\doibase 10.1016/j.cpc.2020.107445} {\bibfield
  {journal} {\bibinfo  {journal} {Comput. Phys. Commun.}\ }\textbf {\bibinfo
  {volume} {256}},\ \bibinfo {pages} {107445} (\bibinfo {year} {2020})},\
  \Eprint {http://arxiv.org/abs/2001.10755} {arXiv:2001.10755 [hep-ph]}
  \BibitemShut {NoStop}%
\bibitem [{\citenamefont {Bondarenko}\ \emph {et~al.}(2023)\citenamefont
  {Bondarenko}, \citenamefont {Dydyshka}, \citenamefont {Kalinovskaya},
  \citenamefont {Sadykov},\ and\ \citenamefont
  {Yermolchyk}}]{Bondarenko:2022mbi}%
  \BibitemOpen
  \bibfield  {author} {\bibinfo {author} {\bibfnamefont {S.}~\bibnamefont
  {Bondarenko}}, \bibinfo {author} {\bibfnamefont {Y.}~\bibnamefont
  {Dydyshka}}, \bibinfo {author} {\bibfnamefont {L.}~\bibnamefont
  {Kalinovskaya}}, \bibinfo {author} {\bibfnamefont {R.}~\bibnamefont
  {Sadykov}}, \ and\ \bibinfo {author} {\bibfnamefont {V.}~\bibnamefont
  {Yermolchyk}},\ }\href {\doibase 10.1016/j.cpc.2022.108646} {\bibfield
  {journal} {\bibinfo  {journal} {Comput. Phys. Commun.}\ }\textbf {\bibinfo
  {volume} {285}},\ \bibinfo {pages} {108646} (\bibinfo {year} {2023})},\
  \Eprint {http://arxiv.org/abs/2207.04332} {arXiv:2207.04332 [hep-ph]}
  \BibitemShut {NoStop}%
\bibitem [{\citenamefont {Banerjee}\ \emph
  {et~al.}(2020{\natexlab{a}})\citenamefont {Banerjee} \emph
  {et~al.}}]{Banerjee:2020tdt}%
  \BibitemOpen
  \bibfield  {author} {\bibinfo {author} {\bibfnamefont {P.}~\bibnamefont
  {Banerjee}} \emph {et~al.},\ }\href {\doibase 10.1140/epjc/s10052-020-8138-9}
  {\bibfield  {journal} {\bibinfo  {journal} {Eur. Phys. J. C}\ }\textbf
  {\bibinfo {volume} {80}},\ \bibinfo {pages} {591} (\bibinfo {year}
  {2020}{\natexlab{a}})},\ \Eprint {http://arxiv.org/abs/2004.13663}
  {arXiv:2004.13663 [hep-ph]} \BibitemShut {NoStop}%
\bibitem [{\citenamefont {Banerjee}\ \emph
  {et~al.}(2020{\natexlab{b}})\citenamefont {Banerjee}, \citenamefont {Engel},
  \citenamefont {Signer},\ and\ \citenamefont {Ulrich}}]{Banerjee:2020rww}%
  \BibitemOpen
  \bibfield  {author} {\bibinfo {author} {\bibfnamefont {P.}~\bibnamefont
  {Banerjee}}, \bibinfo {author} {\bibfnamefont {T.}~\bibnamefont {Engel}},
  \bibinfo {author} {\bibfnamefont {A.}~\bibnamefont {Signer}}, \ and\ \bibinfo
  {author} {\bibfnamefont {Y.}~\bibnamefont {Ulrich}},\ }\href {\doibase
  10.21468/SciPostPhys.9.2.027} {\bibfield  {journal} {\bibinfo  {journal}
  {SciPost Phys.}\ }\textbf {\bibinfo {volume} {9}},\ \bibinfo {pages} {027}
  (\bibinfo {year} {2020}{\natexlab{b}})},\ \Eprint
  {http://arxiv.org/abs/2007.01654} {arXiv:2007.01654 [hep-ph]} \BibitemShut
  {NoStop}%
\bibitem [{\citenamefont {Banerjee}\ \emph {et~al.}(2021)\citenamefont
  {Banerjee}, \citenamefont {Engel}, \citenamefont {Schalch}, \citenamefont
  {Signer},\ and\ \citenamefont {Ulrich}}]{Banerjee:2021mty}%
  \BibitemOpen
  \bibfield  {author} {\bibinfo {author} {\bibfnamefont {P.}~\bibnamefont
  {Banerjee}}, \bibinfo {author} {\bibfnamefont {T.}~\bibnamefont {Engel}},
  \bibinfo {author} {\bibfnamefont {N.}~\bibnamefont {Schalch}}, \bibinfo
  {author} {\bibfnamefont {A.}~\bibnamefont {Signer}}, \ and\ \bibinfo {author}
  {\bibfnamefont {Y.}~\bibnamefont {Ulrich}},\ }\href {\doibase
  10.1016/j.physletb.2021.136547} {\bibfield  {journal} {\bibinfo  {journal}
  {Phys. Lett. B}\ }\textbf {\bibinfo {volume} {820}},\ \bibinfo {pages}
  {136547} (\bibinfo {year} {2021})},\ \Eprint
  {http://arxiv.org/abs/2106.07469} {arXiv:2106.07469 [hep-ph]} \BibitemShut
  {NoStop}%
\bibitem [{\citenamefont {Broggio}\ \emph {et~al.}(2023)\citenamefont {Broggio}
  \emph {et~al.}}]{Broggio:2022htr}%
  \BibitemOpen
  \bibfield  {author} {\bibinfo {author} {\bibfnamefont {A.}~\bibnamefont
  {Broggio}} \emph {et~al.},\ }\href {\doibase 10.1007/JHEP01(2023)112}
  {\bibfield  {journal} {\bibinfo  {journal} {JHEP}\ }\textbf {\bibinfo
  {volume} {01}},\ \bibinfo {pages} {112} (\bibinfo {year} {2023})},\ \Eprint
  {http://arxiv.org/abs/2212.06481} {arXiv:2212.06481 [hep-ph]} \BibitemShut
  {NoStop}%
\bibitem [{\citenamefont {Janot}\ \emph {et~al.}(2024)\citenamefont {Janot},
  \citenamefont {Grojean}, \citenamefont {Zimmermann},\ and\ \citenamefont
  {Benedikt}}]{janot_2024_yr3v6-dgh16}%
  \BibitemOpen
  \bibfield  {author} {\bibinfo {author} {\bibfnamefont {P.}~\bibnamefont
  {Janot}}, \bibinfo {author} {\bibfnamefont {C.}~\bibnamefont {Grojean}},
  \bibinfo {author} {\bibfnamefont {F.}~\bibnamefont {Zimmermann}}, \ and\
  \bibinfo {author} {\bibfnamefont {M.}~\bibnamefont {Benedikt}},\ }\href
  {\doibase 10.17181/yr3v6-dgh16} {\emph {\bibinfo {title} {Integrated
  Luminosities and Sequence of Events for the FCC Feasibility Study Report}}}
  (\bibinfo {year} {2024})\BibitemShut {NoStop}%
\bibitem [{\citenamefont {Buskulic}\ \emph {et~al.}(1995)\citenamefont
  {Buskulic} \emph {et~al.}}]{ALEPH:1994ayc}%
  \BibitemOpen
  \bibfield  {author} {\bibinfo {author} {\bibfnamefont {D.}~\bibnamefont
  {Buskulic}} \emph {et~al.} (\bibinfo {collaboration} {ALEPH}),\ }\href
  {\doibase 10.1016/0168-9002(95)00138-7} {\bibfield  {journal} {\bibinfo
  {journal} {Nucl. Instrum. Meth. A}\ }\textbf {\bibinfo {volume} {360}},\
  \bibinfo {pages} {481} (\bibinfo {year} {1995})}\BibitemShut {NoStop}%
\bibitem [{\citenamefont {Barate}\ \emph {et~al.}(2000)\citenamefont {Barate}
  \emph {et~al.}}]{ALEPH:1999smx}%
  \BibitemOpen
  \bibfield  {author} {\bibinfo {author} {\bibfnamefont {R.}~\bibnamefont
  {Barate}} \emph {et~al.} (\bibinfo {collaboration} {ALEPH}),\ }\href
  {\doibase 10.1007/s100520000319} {\bibfield  {journal} {\bibinfo  {journal}
  {Eur. Phys. J. C}\ }\textbf {\bibinfo {volume} {14}},\ \bibinfo {pages} {1}
  (\bibinfo {year} {2000})}\BibitemShut {NoStop}%
\bibitem [{\citenamefont {Blondel}\ \emph
  {et~al.}(2019{\natexlab{b}})\citenamefont {Blondel} \emph
  {et~al.}}]{Blondel:2019jmp}%
  \BibitemOpen
  \bibfield  {author} {\bibinfo {author} {\bibfnamefont {A.}~\bibnamefont
  {Blondel}} \emph {et~al.},\ }\href@noop {} {\  (\bibinfo {year}
  {2019}{\natexlab{b}})},\ \Eprint {http://arxiv.org/abs/1909.12245}
  {arXiv:1909.12245 [physics.acc-ph]} \BibitemShut {NoStop}%
\bibitem [{\citenamefont {Webber}\ \emph {et~al.}(2011)\citenamefont {Webber}
  \emph {et~al.}}]{MuLan:2010shf}%
  \BibitemOpen
  \bibfield  {author} {\bibinfo {author} {\bibfnamefont {D.~M.}\ \bibnamefont
  {Webber}} \emph {et~al.} (\bibinfo {collaboration} {MuLan}),\ }\href
  {\doibase 10.1103/PhysRevLett.106.079901} {\bibfield  {journal} {\bibinfo
  {journal} {Phys. Rev. Lett.}\ }\textbf {\bibinfo {volume} {106}},\ \bibinfo
  {pages} {041803} (\bibinfo {year} {2011})},\ \Eprint
  {http://arxiv.org/abs/1010.0991} {arXiv:1010.0991 [hep-ex]} \BibitemShut
  {NoStop}%
\bibitem [{\citenamefont {Jegerlehner}(2017)}]{Jegerlehner:2017gek}%
  \BibitemOpen
  \bibfield  {author} {\bibinfo {author} {\bibfnamefont {F.}~\bibnamefont
  {Jegerlehner}},\ }\href {\doibase 10.1007/978-3-319-63577-4} {\emph {\bibinfo
  {title} {{The Anomalous Magnetic Moment of the Muon}}}},\ Vol.\ \bibinfo
  {volume} {274}\ (\bibinfo  {publisher} {Springer},\ \bibinfo {address}
  {Cham},\ \bibinfo {year} {2017})\BibitemShut {NoStop}%
\bibitem [{\citenamefont {Abbiendi}\ \emph {et~al.}(2006)\citenamefont
  {Abbiendi} \emph {et~al.}}]{OPAL:2005xqs}%
  \BibitemOpen
  \bibfield  {author} {\bibinfo {author} {\bibfnamefont {G.}~\bibnamefont
  {Abbiendi}} \emph {et~al.} (\bibinfo {collaboration} {OPAL}),\ }\href
  {\doibase 10.1140/epjc/s2005-02389-3} {\bibfield  {journal} {\bibinfo
  {journal} {Eur. Phys. J. C}\ }\textbf {\bibinfo {volume} {45}},\ \bibinfo
  {pages} {1} (\bibinfo {year} {2006})},\ \Eprint
  {http://arxiv.org/abs/hep-ex/0505072} {arXiv:hep-ex/0505072} \BibitemShut
  {NoStop}%
\bibitem [{\citenamefont {Achard}\ \emph {et~al.}(2005)\citenamefont {Achard}
  \emph {et~al.}}]{L3:2005tsb}%
  \BibitemOpen
  \bibfield  {author} {\bibinfo {author} {\bibfnamefont {P.}~\bibnamefont
  {Achard}} \emph {et~al.} (\bibinfo {collaboration} {L3}),\ }\href {\doibase
  10.1016/j.physletb.2005.07.052} {\bibfield  {journal} {\bibinfo  {journal}
  {Phys. Lett. B}\ }\textbf {\bibinfo {volume} {623}},\ \bibinfo {pages} {26}
  (\bibinfo {year} {2005})},\ \Eprint {http://arxiv.org/abs/hep-ex/0507078}
  {arXiv:hep-ex/0507078} \BibitemShut {NoStop}%
\bibitem [{\citenamefont {Davier}\ \emph {et~al.}(2017)\citenamefont {Davier},
  \citenamefont {Hoecker}, \citenamefont {Malaescu},\ and\ \citenamefont
  {Zhang}}]{Davier:2017zfy}%
  \BibitemOpen
  \bibfield  {author} {\bibinfo {author} {\bibfnamefont {M.}~\bibnamefont
  {Davier}}, \bibinfo {author} {\bibfnamefont {A.}~\bibnamefont {Hoecker}},
  \bibinfo {author} {\bibfnamefont {B.}~\bibnamefont {Malaescu}}, \ and\
  \bibinfo {author} {\bibfnamefont {Z.}~\bibnamefont {Zhang}},\ }\href
  {\doibase 10.1140/epjc/s10052-017-5161-6} {\bibfield  {journal} {\bibinfo
  {journal} {Eur. Phys. J. C}\ }\textbf {\bibinfo {volume} {77}},\ \bibinfo
  {pages} {827} (\bibinfo {year} {2017})},\ \Eprint
  {http://arxiv.org/abs/1706.09436} {arXiv:1706.09436 [hep-ph]} \BibitemShut
  {NoStop}%
\bibitem [{\citenamefont {Harlander}\ and\ \citenamefont
  {Steinhauser}(2003)}]{Harlander:2002ur}%
  \BibitemOpen
  \bibfield  {author} {\bibinfo {author} {\bibfnamefont {R.~V.}\ \bibnamefont
  {Harlander}}\ and\ \bibinfo {author} {\bibfnamefont {M.}~\bibnamefont
  {Steinhauser}},\ }\href {\doibase 10.1016/S0010-4655(03)00204-2} {\bibfield
  {journal} {\bibinfo  {journal} {Comput. Phys. Commun.}\ }\textbf {\bibinfo
  {volume} {153}},\ \bibinfo {pages} {244} (\bibinfo {year} {2003})},\ \Eprint
  {http://arxiv.org/abs/hep-ph/0212294} {arXiv:hep-ph/0212294} \BibitemShut
  {NoStop}%
\bibitem [{\citenamefont {Workman}\ and\ \citenamefont
  {Others}(2022)}]{Workman:2022ynf}%
  \BibitemOpen
  \bibfield  {author} {\bibinfo {author} {\bibfnamefont {R.~L.}\ \bibnamefont
  {Workman}}\ and\ \bibinfo {author} {\bibnamefont {Others}} (\bibinfo
  {collaboration} {Particle Data Group}),\ }\href {\doibase
  10.1093/ptep/ptac097} {\bibfield  {journal} {\bibinfo  {journal} {PTEP}\
  }\textbf {\bibinfo {volume} {2022}},\ \bibinfo {pages} {083C01} (\bibinfo
  {year} {2022})}\BibitemShut {NoStop}%
\bibitem [{\citenamefont {Einhorn}\ \emph {et~al.}(1981)\citenamefont
  {Einhorn}, \citenamefont {Jones},\ and\ \citenamefont
  {Veltman}}]{EINHORN1981146}%
  \BibitemOpen
  \bibfield  {author} {\bibinfo {author} {\bibfnamefont {M.}~\bibnamefont
  {Einhorn}}, \bibinfo {author} {\bibfnamefont {D.}~\bibnamefont {Jones}}, \
  and\ \bibinfo {author} {\bibfnamefont {M.}~\bibnamefont {Veltman}},\ }\href
  {\doibase https://doi.org/10.1016/0550-3213(81)90292-3} {\bibfield  {journal}
  {\bibinfo  {journal} {Nuclear Physics B}\ }\textbf {\bibinfo {volume}
  {191}},\ \bibinfo {pages} {146} (\bibinfo {year} {1981})}\BibitemShut
  {NoStop}%
\bibitem [{\citenamefont {Chanowitz}\ \emph {et~al.}(1978)\citenamefont
  {Chanowitz}, \citenamefont {Furman},\ and\ \citenamefont
  {Hinchliffe}}]{CHANOWITZ1978285}%
  \BibitemOpen
  \bibfield  {author} {\bibinfo {author} {\bibfnamefont {M.}~\bibnamefont
  {Chanowitz}}, \bibinfo {author} {\bibfnamefont {M.}~\bibnamefont {Furman}}, \
  and\ \bibinfo {author} {\bibfnamefont {I.}~\bibnamefont {Hinchliffe}},\
  }\href {\doibase https://doi.org/10.1016/0370-2693(78)90024-2} {\bibfield
  {journal} {\bibinfo  {journal} {Physics Letters B}\ }\textbf {\bibinfo
  {volume} {78}},\ \bibinfo {pages} {285} (\bibinfo {year} {1978})}\BibitemShut
  {NoStop}%
\bibitem [{\citenamefont {Consoli}\ \emph {et~al.}(1989)\citenamefont
  {Consoli}, \citenamefont {Hollik},\ and\ \citenamefont
  {Jegerlehner}}]{Consoli:1989fg}%
  \BibitemOpen
  \bibfield  {author} {\bibinfo {author} {\bibfnamefont {M.}~\bibnamefont
  {Consoli}}, \bibinfo {author} {\bibfnamefont {W.}~\bibnamefont {Hollik}}, \
  and\ \bibinfo {author} {\bibfnamefont {F.}~\bibnamefont {Jegerlehner}},\
  }\href {\doibase 10.1016/0370-2693(89)91301-4} {\bibfield  {journal}
  {\bibinfo  {journal} {Phys. Lett. B}\ }\textbf {\bibinfo {volume} {227}},\
  \bibinfo {pages} {167} (\bibinfo {year} {1989})}\BibitemShut {NoStop}%
\bibitem [{\citenamefont {Altarelli}\ and\ \citenamefont
  {Barbieri}(1991)}]{Altarelli:1990zd}%
  \BibitemOpen
  \bibfield  {author} {\bibinfo {author} {\bibfnamefont {G.}~\bibnamefont
  {Altarelli}}\ and\ \bibinfo {author} {\bibfnamefont {R.}~\bibnamefont
  {Barbieri}},\ }\href {\doibase 10.1016/0370-2693(91)91378-9} {\bibfield
  {journal} {\bibinfo  {journal} {Phys. Lett. B}\ }\textbf {\bibinfo {volume}
  {253}},\ \bibinfo {pages} {161} (\bibinfo {year} {1991})}\BibitemShut
  {NoStop}%
\bibitem [{\citenamefont {Peskin}\ and\ \citenamefont
  {Takeuchi}(1992)}]{PhysRevD.46.381}%
  \BibitemOpen
  \bibfield  {author} {\bibinfo {author} {\bibfnamefont {M.~E.}\ \bibnamefont
  {Peskin}}\ and\ \bibinfo {author} {\bibfnamefont {T.}~\bibnamefont
  {Takeuchi}},\ }\href {\doibase 10.1103/PhysRevD.46.381} {\bibfield  {journal}
  {\bibinfo  {journal} {Phys. Rev. D}\ }\textbf {\bibinfo {volume} {46}},\
  \bibinfo {pages} {381} (\bibinfo {year} {1992})}\BibitemShut {NoStop}%
\bibitem [{\citenamefont {Barbieri}\ \emph {et~al.}(2004)\citenamefont
  {Barbieri}, \citenamefont {Pomarol}, \citenamefont {Rattazzi},\ and\
  \citenamefont {Strumia}}]{Barbieri:2004qk}%
  \BibitemOpen
  \bibfield  {author} {\bibinfo {author} {\bibfnamefont {R.}~\bibnamefont
  {Barbieri}}, \bibinfo {author} {\bibfnamefont {A.}~\bibnamefont {Pomarol}},
  \bibinfo {author} {\bibfnamefont {R.}~\bibnamefont {Rattazzi}}, \ and\
  \bibinfo {author} {\bibfnamefont {A.}~\bibnamefont {Strumia}},\ }\href
  {\doibase 10.1016/j.nuclphysb.2004.10.014} {\bibfield  {journal} {\bibinfo
  {journal} {Nucl. Phys. B}\ }\textbf {\bibinfo {volume} {703}},\ \bibinfo
  {pages} {127} (\bibinfo {year} {2004})},\ \Eprint
  {http://arxiv.org/abs/hep-ph/0405040} {arXiv:hep-ph/0405040} \BibitemShut
  {NoStop}%
\bibitem [{\citenamefont {Hayrapetyan}\ \emph {et~al.}(2024)\citenamefont
  {Hayrapetyan} \emph {et~al.}}]{ATLAS:2024dxp}%
  \BibitemOpen
  \bibfield  {author} {\bibinfo {author} {\bibfnamefont {A.}~\bibnamefont
  {Hayrapetyan}} \emph {et~al.} (\bibinfo {collaboration} {ATLAS, CMS}),\
  }\href {\doibase 10.1103/PhysRevLett.132.261902} {\bibfield  {journal}
  {\bibinfo  {journal} {Phys. Rev. Lett.}\ }\textbf {\bibinfo {volume} {132}},\
  \bibinfo {pages} {261902} (\bibinfo {year} {2024})},\ \Eprint
  {http://arxiv.org/abs/2402.08713} {arXiv:2402.08713 [hep-ex]} \BibitemShut
  {NoStop}%
\bibitem [{\citenamefont {Nason}(2019)}]{Nason:2017cxd}%
  \BibitemOpen
  \bibfield  {author} {\bibinfo {author} {\bibfnamefont {P.}~\bibnamefont
  {Nason}},\ }\enquote {\bibinfo {title} {{The Top Mass in Hadronic
  Collisions}},}\ in\ \href {\doibase 10.1142/9789813238053_0008} {\emph
  {\bibinfo {booktitle} {{From My Vast Repertoire ...}: {Guido Altarelli's
  Legacy}}}},\ \bibinfo {editor} {edited by\ \bibinfo {editor} {\bibfnamefont
  {A.}~\bibnamefont {Levy}}, \bibinfo {editor} {\bibfnamefont {S.}~\bibnamefont
  {Forte}}, \ and\ \bibinfo {editor} {\bibfnamefont {G.}~\bibnamefont
  {Ridolfi}}}\ (\bibinfo {year} {2019})\ pp.\ \bibinfo {pages} {123--151},\
  \Eprint {http://arxiv.org/abs/1712.02796} {arXiv:1712.02796 [hep-ph]}
  \BibitemShut {NoStop}%
\bibitem [{\citenamefont {Azzi}\ \emph {et~al.}(2019)\citenamefont {Azzi} \emph
  {et~al.}}]{Azzi:2019yne}%
  \BibitemOpen
  \bibfield  {author} {\bibinfo {author} {\bibfnamefont {P.}~\bibnamefont
  {Azzi}} \emph {et~al.},\ }\href {\doibase 10.23731/CYRM-2019-007.1}
  {\bibfield  {journal} {\bibinfo  {journal} {CERN Yellow Rep. Monogr.}\
  }\textbf {\bibinfo {volume} {7}},\ \bibinfo {pages} {1} (\bibinfo {year}
  {2019})},\ \Eprint {http://arxiv.org/abs/1902.04070} {arXiv:1902.04070
  [hep-ph]} \BibitemShut {NoStop}%
\bibitem [{\citenamefont {Hoang}(2020)}]{Hoang:2020iah}%
  \BibitemOpen
  \bibfield  {author} {\bibinfo {author} {\bibfnamefont {A.~H.}\ \bibnamefont
  {Hoang}},\ }\href {\doibase 10.1146/annurev-nucl-101918-023530} {\bibfield
  {journal} {\bibinfo  {journal} {Ann. Rev. Nucl. Part. Sci.}\ }\textbf
  {\bibinfo {volume} {70}},\ \bibinfo {pages} {225} (\bibinfo {year} {2020})},\
  \Eprint {http://arxiv.org/abs/2004.12915} {arXiv:2004.12915 [hep-ph]}
  \BibitemShut {NoStop}%
\bibitem [{\citenamefont {Dehnadi}\ \emph {et~al.}(2023)\citenamefont
  {Dehnadi}, \citenamefont {Hoang}, \citenamefont {Jin},\ and\ \citenamefont
  {Mateu}}]{Dehnadi:2023msm}%
  \BibitemOpen
  \bibfield  {author} {\bibinfo {author} {\bibfnamefont {B.}~\bibnamefont
  {Dehnadi}}, \bibinfo {author} {\bibfnamefont {A.~H.}\ \bibnamefont {Hoang}},
  \bibinfo {author} {\bibfnamefont {O.~L.}\ \bibnamefont {Jin}}, \ and\
  \bibinfo {author} {\bibfnamefont {V.}~\bibnamefont {Mateu}},\ }\href
  {\doibase 10.1007/JHEP12(2023)065} {\bibfield  {journal} {\bibinfo  {journal}
  {JHEP}\ }\textbf {\bibinfo {volume} {12}},\ \bibinfo {pages} {065} (\bibinfo
  {year} {2023})},\ \Eprint {http://arxiv.org/abs/2309.00547} {arXiv:2309.00547
  [hep-ph]} \BibitemShut {NoStop}%
\bibitem [{\citenamefont {Tumasyan}\ \emph
  {et~al.}(2023{\natexlab{a}})\citenamefont {Tumasyan} \emph
  {et~al.}}]{CMS:2023ebf}%
  \BibitemOpen
  \bibfield  {author} {\bibinfo {author} {\bibfnamefont {A.}~\bibnamefont
  {Tumasyan}} \emph {et~al.} (\bibinfo {collaboration} {CMS}),\ }\href
  {\doibase 10.1140/epjc/s10052-023-12050-4} {\bibfield  {journal} {\bibinfo
  {journal} {Eur. Phys. J. C}\ }\textbf {\bibinfo {volume} {83}},\ \bibinfo
  {pages} {963} (\bibinfo {year} {2023}{\natexlab{a}})},\ \Eprint
  {http://arxiv.org/abs/2302.01967} {arXiv:2302.01967 [hep-ex]} \BibitemShut
  {NoStop}%
\bibitem [{\citenamefont {Aad}\ \emph {et~al.}(2023)\citenamefont {Aad} \emph
  {et~al.}}]{ATLAS:2022jbw}%
  \BibitemOpen
  \bibfield  {author} {\bibinfo {author} {\bibfnamefont {G.}~\bibnamefont
  {Aad}} \emph {et~al.} (\bibinfo {collaboration} {ATLAS}),\ }\href {\doibase
  10.1007/JHEP06(2023)019} {\bibfield  {journal} {\bibinfo  {journal} {JHEP}\
  }\textbf {\bibinfo {volume} {06}},\ \bibinfo {pages} {019} (\bibinfo {year}
  {2023})},\ \Eprint {http://arxiv.org/abs/2209.00583} {arXiv:2209.00583
  [hep-ex]} \BibitemShut {NoStop}%
\bibitem [{CDF(2016)}]{CDF:2016vzt}%
  \BibitemOpen
  \href@noop {} {\  (\bibinfo {year} {2016})},\ \Eprint
  {http://arxiv.org/abs/1608.01881} {arXiv:1608.01881 [hep-ex]} \BibitemShut
  {NoStop}%
\bibitem [{\citenamefont {Tumasyan}\ \emph
  {et~al.}(2023{\natexlab{b}})\citenamefont {Tumasyan} \emph
  {et~al.}}]{CMS:2022emx}%
  \BibitemOpen
  \bibfield  {author} {\bibinfo {author} {\bibfnamefont {A.}~\bibnamefont
  {Tumasyan}} \emph {et~al.} (\bibinfo {collaboration} {CMS}),\ }\href
  {\doibase 10.1007/JHEP07(2023)077} {\bibfield  {journal} {\bibinfo  {journal}
  {JHEP}\ }\textbf {\bibinfo {volume} {07}},\ \bibinfo {pages} {077} (\bibinfo
  {year} {2023}{\natexlab{b}})},\ \Eprint {http://arxiv.org/abs/2207.02270}
  {arXiv:2207.02270 [hep-ex]} \BibitemShut {NoStop}%
\bibitem [{\citenamefont {Dubovyk}\ \emph {et~al.}(2018)\citenamefont
  {Dubovyk}, \citenamefont {Freitas}, \citenamefont {Gluza}, \citenamefont
  {Riemann},\ and\ \citenamefont {Usovitsch}}]{Dubovyk:2018rlg}%
  \BibitemOpen
  \bibfield  {author} {\bibinfo {author} {\bibfnamefont {I.}~\bibnamefont
  {Dubovyk}}, \bibinfo {author} {\bibfnamefont {A.}~\bibnamefont {Freitas}},
  \bibinfo {author} {\bibfnamefont {J.}~\bibnamefont {Gluza}}, \bibinfo
  {author} {\bibfnamefont {T.}~\bibnamefont {Riemann}}, \ and\ \bibinfo
  {author} {\bibfnamefont {J.}~\bibnamefont {Usovitsch}},\ }\href {\doibase
  10.1016/j.physletb.2018.06.037} {\bibfield  {journal} {\bibinfo  {journal}
  {Phys. Lett. B}\ }\textbf {\bibinfo {volume} {783}},\ \bibinfo {pages} {86}
  (\bibinfo {year} {2018})},\ \Eprint {http://arxiv.org/abs/1804.10236}
  {arXiv:1804.10236 [hep-ph]} \BibitemShut {NoStop}%
\bibitem [{\citenamefont {de~Florian}\ \emph {et~al.}(2016)\citenamefont
  {de~Florian} \emph {et~al.}}]{LHCHiggsCrossSectionWorkingGroup:2016ypw}%
  \BibitemOpen
  \bibfield  {author} {\bibinfo {author} {\bibfnamefont {D.}~\bibnamefont
  {de~Florian}} \emph {et~al.} (\bibinfo {collaboration} {LHC Higgs Cross
  Section Working Group}),\ }\href {\doibase 10.23731/CYRM-2017-002} {\ \textbf
  {\bibinfo {volume} {2/2017}} (\bibinfo {year} {2016}),\
  10.23731/CYRM-2017-002},\ \Eprint {http://arxiv.org/abs/1610.07922}
  {arXiv:1610.07922 [hep-ph]} \BibitemShut {NoStop}%
\bibitem [{\citenamefont {de~Blas}\ \emph {et~al.}(2017)\citenamefont
  {de~Blas}, \citenamefont {Ciuchini}, \citenamefont {Franco}, \citenamefont
  {Mishima}, \citenamefont {Pierini}, \citenamefont {Reina},\ and\
  \citenamefont {Silvestrini}}]{deBlas:2016nqo}%
  \BibitemOpen
  \bibfield  {author} {\bibinfo {author} {\bibfnamefont {J.}~\bibnamefont
  {de~Blas}}, \bibinfo {author} {\bibfnamefont {M.}~\bibnamefont {Ciuchini}},
  \bibinfo {author} {\bibfnamefont {E.}~\bibnamefont {Franco}}, \bibinfo
  {author} {\bibfnamefont {S.}~\bibnamefont {Mishima}}, \bibinfo {author}
  {\bibfnamefont {M.}~\bibnamefont {Pierini}}, \bibinfo {author} {\bibfnamefont
  {L.}~\bibnamefont {Reina}}, \ and\ \bibinfo {author} {\bibfnamefont
  {L.}~\bibnamefont {Silvestrini}},\ }\href {\doibase 10.22323/1.282.0690}
  {\bibfield  {journal} {\bibinfo  {journal} {PoS}\ }\textbf {\bibinfo {volume}
  {ICHEP2016}},\ \bibinfo {pages} {690} (\bibinfo {year} {2017})},\ \Eprint
  {http://arxiv.org/abs/1611.05354} {arXiv:1611.05354 [hep-ph]} \BibitemShut
  {NoStop}%
\bibitem [{\citenamefont {de~Blas}\ \emph {et~al.}(2022)\citenamefont
  {de~Blas}, \citenamefont {Du}, \citenamefont {Grojean}, \citenamefont {Gu},
  \citenamefont {Miralles}, \citenamefont {Peskin}, \citenamefont {Tian},
  \citenamefont {Vos},\ and\ \citenamefont {Vryonidou}}]{deBlas:2022ofj}%
  \BibitemOpen
  \bibfield  {author} {\bibinfo {author} {\bibfnamefont {J.}~\bibnamefont
  {de~Blas}}, \bibinfo {author} {\bibfnamefont {Y.}~\bibnamefont {Du}},
  \bibinfo {author} {\bibfnamefont {C.}~\bibnamefont {Grojean}}, \bibinfo
  {author} {\bibfnamefont {J.}~\bibnamefont {Gu}}, \bibinfo {author}
  {\bibfnamefont {V.}~\bibnamefont {Miralles}}, \bibinfo {author}
  {\bibfnamefont {M.~E.}\ \bibnamefont {Peskin}}, \bibinfo {author}
  {\bibfnamefont {J.}~\bibnamefont {Tian}}, \bibinfo {author} {\bibfnamefont
  {M.}~\bibnamefont {Vos}}, \ and\ \bibinfo {author} {\bibfnamefont
  {E.}~\bibnamefont {Vryonidou}},\ }in\ \href@noop {} {\emph {\bibinfo
  {booktitle} {{Snowmass 2021}}}}\ (\bibinfo {year} {2022})\ \Eprint
  {http://arxiv.org/abs/2206.08326} {arXiv:2206.08326 [hep-ph]} \BibitemShut
  {NoStop}%
\end{thebibliography}%

\end{document}